\documentclass[transmag]{IEEEtran}

\usepackage{amsmath,amsthm,amsbsy,amssymb,amstext,amsfonts} % Good for text in math mode.
\usepackage{algorithm}
\usepackage{algpseudocode}

\usepackage[english]{babel}
\usepackage{bm}
\usepackage{bbold}

\usepackage[font=footnotesize,justification=justified,format=plain]{caption}
\usepackage[compress]{cite}
\usepackage{color}

\usepackage{graphicx}

\usepackage{hyperref}

\usepackage{import} 
\usepackage{latexsym}

\usepackage{listings}

\usepackage{overpic} 

\usepackage{setspace}
\usepackage[labelformat=simple,font=footnotesize,justification=justified,format=plain]{subcaption}

\usepackage{textcomp}

\usepackage{url}

\usepackage{xcolor}

\newcommand{\mbtwoD}{\mbox{2-D}}
\newcommand{\mbCf}{\mbox{Cell-free}}
\newcommand{\mbcf}{\mbox{cell-free}}
\newcommand{\mbcv}{\mbox{cross-validation}}
\newcommand{\mbol}{\mbox{off-line}}
\newcommand{\mbdml}{\mbox{Dask-ML}}
\newcommand{\mbfl}{\mbox{for-loop}}
\newcommand{\mbfls}{\mbox{for-loops}}
\newcommand{\mbio}{\mbox{input-output}}
\newcommand{\mbkf}{\mbox{K-factor}}
\newcommand{\mblos}{\mbox{line-of-sight}}
\newcommand{\mbln}{\mbox{log-normal}}
\newcommand{\mblc}{\mbox{low-cost}}
\newcommand{\mblr}{\mbox{low-rank}}
\newcommand{\mblrs}{\mbox{low-ranks}}
\newcommand{\mbmd}{\mbox{macro-diversity}}
\newcommand{\mbmmw}{\mbox{mm-wave}}

\newcommand{\mbma}{\mbox{multi-antenna}}
\newcommand{\mbmc}{\mbox{multi-class}}
\newcommand{\mbmgbps}{\mbox{multi-Gbps}}
\newcommand{\mbml}{\mbox{multi-label}}
\newcommand{\mbmo}{\mbox{multi-output}}
\newcommand{\mbmp}{\mbox{multi-path}}

\newcommand{\mbms}{\mbox{multi-stream}}
\newcommand{\mbot}{\mbox{one-time}}
\newcommand{\mbpu}{\mbox{per-user}}
\newcommand{\mbSV}{\mbox{Saleh-Valenzuela}}
\newcommand{\mbsc}{\mbox{semi-centralized}}
\newcommand{\mbsl}{\mbox{single-label}}
\newcommand{\mbslin}{\mbox{semilinear}}

\newcommand{\mbso}{\mbox{single-output}}
\newcommand{\mbsr}{\mbox{sum-rate}}
\newcommand{\mbsrs}{\mbox{sum-rates}}
\newcommand{\mbto}{\mbox{trade-off}}
\newcommand{\mbts}{\mbox{tree-structured}}
\newcommand{\mbtsta}{\mbox{two-stage}}
\newcommand{\mbup}{\mbox{unit-power}}
\newcommand{\mbwc}{\mbox{well-constructed}}
\newcommand{\mbwe}{\mbox{well-equipped}}

\newcommand{\mclU}{\mathcal{\underline{U}}}

\newcommand{\mbfa}{\mathbf{a}}
\newcommand{\mbfb}{\mathbf{b}}
\newcommand{\mbfD}{\mathbf{D}}
\newcommand{\mbfh}{\mathbf{h}}
\newcommand{\mbfH}{\mathbf{H}}
\newcommand{\mbfI}{\mathbf{I}}
\newcommand{\mbfL}{\mathbf{L}}
\newcommand{\mbfu}{\mathbf{u}}
\newcommand{\mbflu}{\mathbf{\underline{u}}}
\newcommand{\mbftu}{\tilde{\mathbf{u}}}
\newcommand{\mbflU}{\mathbf{\underline{U}}}
\newcommand{\mbfU}{\mathbf{U}}
\newcommand{\mbfv}{\mathbf{v}}
\newcommand{\mbfV}{\mathbf{V}}

\newcommand{\sikl}{_{kl}}
\newcommand{\siil}{_{il}}
\newcommand{\sijl}{_{jl}}
\newcommand{\sijm}{_{jm}}
\newcommand{\sivec}{_{\mathrm{vec}}}
\newcommand{\ssth}{^{\text{th}}}
\newcommand{\ssnd}{^{\text{nd}}}

\newcommand{\tcomb}{\text{comb}}
\newcommand{\tDL}{\text{DL}}
\newcommand{\tk}{\text{k}}
\newcommand{\tK}{\text{K}}
\newcommand{\tL}{\text{L}}
\newcommand{\tM}{\text{M}}
\newcommand{\tm}{{\text{max}}}
\newcommand{\tmin}{{\text{min}}}
\newcommand{\topt}{\text{opt}}
\newcommand{\trf}{\text{rf}}
\newcommand{\tR}{\text{R}}
\newcommand{\tSINR}{\text{SINR}}

\newcommand{\tst}{\text{s.t.}}
\newcommand{\ts}{\text{sum}}
\newcommand{\tT}{\text{T}}

\definecolor{MyBlue}{RGB}{68,114,196}
\def\BibTeX{{\rm B\kern-.05em{\sc i\kern-.025em b}\kern-.08em T\kern-.1667em\lower.7ex\hbox{E}\kern-.125emX}}
\begin{document}

\title{Joint Analog Beam Selection and Digital Beamforming in Millimeter Wave Cell-Free Massive MIMO Systems}

\author{Cenk M. Yetis*,~\IEEEmembership{Member,~IEEE}, Emil Bj\"{o}rnson,~\IEEEmembership{Senior Member,~IEEE}, and Pontus Giselsson
\thanks{C. M. Yetis (corresponding author) and P. Giselsson are with the Department of Automatic Control, Lund University, 22100, Lund, Sweden (cenkmyetis@ieee.org; \mbox{pontus.giselsson@control.lth.se}).} 
\thanks{E. Bj\"{o}rnson is with the Department of Electrical Engineering (ISY), Link\"{o}ping University, 58183, Link\"{o}ping, Sweden (emil.bjornson@liu.se).}}

\IEEEtitleabstractindextext{\begin{abstract}
\textbf{Cell-free massive MIMO systems consist of many distributed access points with simple components that jointly serve the users. In millimeter wave bands, only a limited set of predetermined beams can be supported. In a network that consolidates these technologies, downlink analog beam selection stands as a challenging task for the network $\mbsr$ maximization. \mbox{Low-cost} digital filters can improve the network $\mbsr$ further. In this work, we propose $\mblc$ joint designs of analog beam selection and digital filters. The proposed joint designs achieve significantly higher $\mbsrs$ than the disjoint design benchmark.  Supervised machine learning (ML) algorithms can efficiently approximate the $\mbio$ mapping functions of the beam selection decisions of the joint designs with low computational complexities. Since the training of ML algorithms is performed $\mbol$, we propose a $\mbwc$ joint design that combines multiple initializations, iterations, and selection features, as well as beam conflict control, i.e., the same beam cannot be used for multiple users. The numerical results indicate that ML algorithms can retain ${99\text{-}100\%}$ of the original sum-rate results achieved by the proposed $\mbwc$ designs.}
\end{abstract}

}

\maketitle

\section{INTRODUCTION}\label{sec:Introduction}

\IEEEPARstart{C}{ell-free} massive MIMO (mMIMO) networks thrive on the idea of jointly and coherently serving a proportionally small number of users by a large number of simple $\mbma$ access points (APs). Compared to cellular mMIMO networks, $\mbcf$ networks can provide a more uniform service performance for the users in the network since the antennas are distributed. For example, the $95\%\text{-likely}$ $\mbpu$ spectral and energy efficiencies of $\mbcf$ networks are five and ten times higher than cellular networks, respectively \cite{7827017}. 

Millimeter wave ($\mbmmw$) communications can achieve $\mbmgbps$ data rates by exploiting underutilized wide bandwidths in the $\mbmmw$ spectrum. Therefore, the consolidation of $\mbmmw$ communications and $\mbcf$ networks is a promising direction for the next generation wireless networks \cite{8676377}. In particular, the $\mbmd$ achieved by having many distributed APs compensate for the spotty coverage that otherwise limits the practical use of $\mbmmw$ spectrum. The hardware complexity of the AP is critical to reduce the costs of deployment and power consumption of APs that are deployed in large numbers in the $\mbcf$ network. Thus, low complexity hybrid analog and digital beamforming designs need to be adopted.

In this work, we propose $\mblc$ joint design algorithms for  analog beam selection and digital beamforming in the downlink transmission of a $\mbmmw$ $\mbcf$ mMIMO system consisting of multiple antenna APs and single antenna users. We assume APs are equipped with uniform linear arrays (ULAs) and the number of radio frequency (RF) chains  at an AP is equal to the number of transmitted streams. The analog beam selection process does not rely on an external aid, e.g., location information of the users, but only on the $\mbsr$ metrics of the users. Furthermore, we incorporate multiple initializations, iterations, and selection features, as well as beam conflict control (BCC), i.e., a selected beam for a user cannot be reselected for another user. We refer to a joint design with multiple features as a $\mbwc$ design. The proposed joint design solutions achieve significant network $\mbsr$ gains compared to the naive disjoint design of analog beam selection, which is based on the direct link (DL) power metrics of the users, and digital precoder. The disjoint approach first completes the analog beam selection and then completes the digital precoder design. On the other hand, the proposed joint approach iteratively updates the beam selections and digital precoder designs until convergence.

Finally, for analog beam selection, we propose machine learning (ML) algorithms that are trained $\mbol$ by the proposed $\mbwc$ designs. Online beam selection by an ML algorithm which is succeeded by digital precoder designs are \mbox{one-time-only}, i.e., no \mbox{for-loops}, and this approach can mirror our proposed $\mbwc$ designs. The proposed ML based approach can achieve ${99\text{-}100\%}$ of the original sum-rate results achieved by the $\mbwc$ designs.

\section{Related Works and Contributions}

Beamforming designs with digital precoders for microwave communications in $\mbcf$ networks are investigated in \cite{7917284,9069486}. ZF, minimum mean squared error (MMSE), \mbox{maximum-ratio} transmitter/combiner (MRT/MRC) digital filters can be preferred in the baseband processing of $\mbmmw$ networks \cite{8322248,8630933,Han2018} for their low computational complexities. 

In $\mbmmw$ networks, proper selections of beams are achieved by transmitting the candidate beams and measuring their performances by a network metric. This process is known as beam training. The conventional beam training is time and power inefficient since it sweeps all beam directions exhaustively. The hierarchical codebook approach lowers the search complexity by implementing a $\mbts$ codebook design \cite{7604098}.  Nevertheless, the complexity of hierarchical search is still high \cite{7914759}.

In \cite{7134756,7790909} compressive sensing is implemented to lower the beam selection complexity. By exploiting the spatial sparsity of $\mbmmw$ channels,  beam selection is achieved implicitly in the \mbox{reduced-dimensional} beamspace CSI domain. In \cite{6587076}, the computational complexity is reduced by combining probabilistic framework based simulated annealing and the $\mbtwoD$ numerical method based Rosenbrock search procedure. In \cite{8760516}, a new antenna array that can adjust the gain and phase of each antenna is used to support $\mbms$ transmission with fewer number of RF chains. In \cite{8599167}, two partial beam training strategies are proposed with reduced computational complexities.  In \cite{8753525}, the computational complexity is reduced by a $\mbtsta$ beam training procedure. In \cite{Wang2019a}, location assistance is proposed to eliminate the need for beam sweeping at the cost of increased power consumption due to continuous global positioning system (GPS) connectivity. In \cite{Song2018}, a \mbox{multi-user} scalable  and channel variation robust beam searching method is proposed. In \cite{Gao2017a}, a branch-and-bound based beam searching algorithm is proposed for a hybrid design with a subarray architecture. 

Ideally, the beam searching complexity should be linear in the network parameters, i.e., the number of APs, users, and the number of antennas at APs, to achieve scalability with the network size. In this work, given a beamforming codebook, we propose two beam searching algorithms, namely $\mbslin$ and linear search algorithms. The search complexity of the former is exponential with respect to the number of users but linear with the number of APs whereas the search complexity of the latter is linear both with the number of APs and users. 

The assignment of the same beam to multiple users results in beam conflict in $\mbmmw$ networks. Beam conflict causes effective channels to have $\mblrs$ which can significantly reduce the $\mbsr$ of the network \cite{8599167}. In this work, we propose BCC to entirely eliminate the beam conflict in the network. In $\mbcf$ networks, BCC is more challenging since an assigned beam to a user can be decided to be used by another AP for the same user as well. On the  contrary, for instance, in interference networks, where a transmitter sends a different stream to each user, a different beam is assigned between each AP and a user. Thus, the assigned beam can be immediately removed from the common codebook of the network. As a solution to this challenge in $\mbcf$ networks, in this work, we propose to keep a codebook log for each user which is updated and announced in the network after each beam assignment. BCC implementation can also reduce the simulation duration due to the reduced number of beam combinations. In this work, we also propose BCC initializations, i.e., random initializations that satisfy BCC, for the algorithms without the BCC implementation. 

Another major challenge in $\mbcf$ networks is that an AP needs to equip as many RF chains as the number of users since APs jointly serve all users . However, there can be opportunities in the varying channel conditions to shut off RF chains to gain significant power savings at low $\mbsr$ losses. In this work, we also propose to adaptively shut off the RF chains that leads to $30\%$ power gain at the cost of $5\%$ $\mbsr$ loss on average.

As the network size grows, even low complexity beam selection algorithms can be too costly. Next, we propose fast and efficient supervised ML algorithms which can be trained by the proposed $\mbwc$ algorithms. In particular, three  ML algorithms motivate us, the support vector machine (SVM),  multi-layer perceptron (MLP), and random forest (RFt) algorithms. BCC introduces patterns on the outputs of the mapping functions in the beam selection problems. Hence, the labels that are input to ML algorithms become highly correlated. Classifier chains are proven to be effective in exploiting the correlations in $\mbml$ ML problems \cite{101007}. Our numerical results indicate that the RFt algorithm with the classifier chains can retain $99\%$ of the original $\mbsr$ results that are achieved by the beam selection algorithms. This is a sharp contrast to $63\%$ achieved without the classifier chains.

Since the training of ML algorithms is carried out $\mbol$, in this work, a $\mbwc$ joint analog beam selection and digital beamforming algorithm is proposed. In particular, we propose a joint design with multiple initializations and iterations. Furthermore, due to BCC, the assignments of beams to the users in the earlier search segments impinge the  assignments in the later search segments. In addition to multiple initializations and iterations, we integrate a selection feature to the proposed joint design as well. Numerical results indicate that the linear search algorithm with the selection feature can achieve higher $\mbsr$ at a lower complexity even than the $\mbslin$ algorithm that has no selection feature.

The rest of the paper is organized as follows. In Section \ref{sec:SystemModel}, we introduce the system model of $\mbmmw$ $\mbcf$ mMIMO networks. In Section \ref{sec:ProblemFormulation}, the problem formulation of joint design is introduced. In Section \ref{sec:BeamSelection}, the proposed low complexity beam searching algorithms are introduced, and the challenge of BCC in $\mbcf$ networks is addressed and a solution is proposed. In Section \ref{sec:SupervisedML}, the solution to $\mbmo$ classification problem under BCC is provided and ML implementation is outlined. In Section \ref{sec:NumericalResults}, the numerical results of joint design and ML algorithms are presented, and finally, the paper is concluded in Section \ref{sec:Conclusion}. 

\textit{Notations:} Throughout the paper, $(.)^H$ and $(.)^{-1}$ denote the conjugate transpose and the inverse operations of a matrix, respectively. $\|.\|_2$ denotes the $L_2$ vector norm operator.  ${\mathcal{CN}(0,x)}$ denotes the complex Gaussian distribution with zero mean and variance $x$. $E\{.\}$ and $|.|$ are the expectation and absolute value operators, respectively. Finally, $!$ denotes the factorial operator.

\section{System Model} \label{sec:SystemModel}
\begin{figure}[!t]
\centering
\begin{subfigure}[t]{0.5\textwidth}
\centering
\begin{overpic}[scale=0.8] {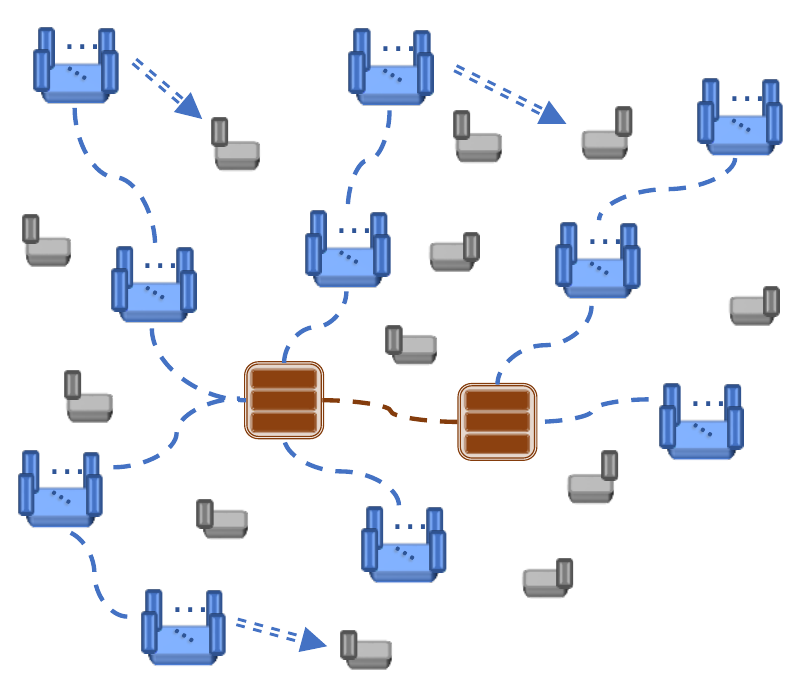}
\put(5,85){\color{MyBlue}\footnotesize AP $1$} \put(45,85){\color{MyBlue}\footnotesize AP $l$}
\put(18,15){\color{MyBlue}\footnotesize AP $L$}
\put(27,72){\color{gray}\footnotesize User $1$} \put(73,75){\color{gray}\footnotesize User $k$}
\put(45,8){\color{gray}\footnotesize User $K$}
\put(19,79){$\mbfh_{11}$}
\put(60,79){$\mbfh\sikl$}
\put(30,10){$\mbfh_{KL}$}
\end{overpic}
\caption{Cell-free network model.}  
\label{Fig:CFModel}
\end{subfigure}

\begin{subfigure}[t]{0.5\textwidth}
\centering
\begin{overpic}[scale=1.25] {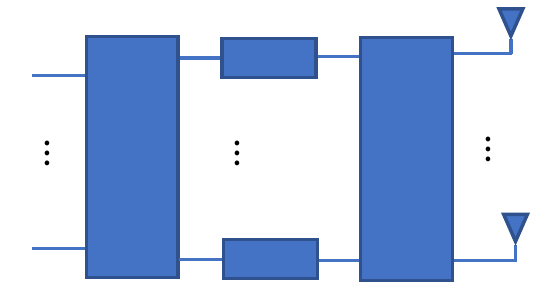}
\put(1,39){\large$s_1$} \put(0,8){\large$s_K$}
\put(45,24){\large$M_\trf$} \put(91,24){\large$M$}
\put(20,28){\large$\mbfV_l$} \put(70,28){\large$\mbfU_l$}
\put(18,20){\color{white}\footnotesize\textbf{Digital}}
\put(17,16){\color{white}\footnotesize\textbf{precoder}}
\put(68,20){\color{white}\footnotesize\textbf{Analog}}
\put(67,16){\color{white}\footnotesize\textbf{precoder}}
\put(42,41.5){\color{white}\footnotesize\textbf{RF chain}}
\put(42,5){\color{white}\footnotesize\textbf{RF chain}}
\end{overpic}
\caption{AP model.}  
\label{Fig:APModel}
\end{subfigure}

\caption{Cell-free and AP models.}  
\label{Fig:Models}
\end{figure}
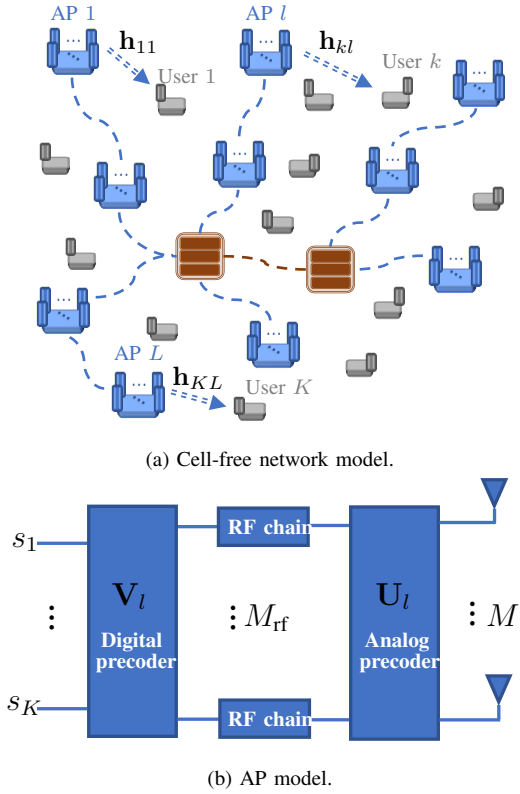

We consider the downlink transmission of a $\mbmmw$ $\mbcf$ mMIMO system consisting of $L$ APs and $K$ users. Each AP has $M$ antennas, while each user has a single antenna. In contrast to $\mbcf$ systems built for \mbox{sub-$6$} GHz systems, the number AP antennas can be very large in $\mbmmw$ systems. The channel between AP $l$ and user $k$ is ${\mbfh\sikl\in\mathbb{C}^M}$, and it is assumed to be frequency-flat. The system and AP models are shown in Fig. \ref{Fig:Models}.

\subsection{Received Signal} \label{subsec:ReceivedSignal}
Consider a hybrid beamforming case where $M_\trf$ is the number of radio frequency chains at an AP and, in general, ${M\geq M_\trf\geq K}$. The radio frequency (analog) and baseband (digital) precoders at AP $l$, ${\forall l\in\mathcal{L}\triangleq\{1,\ldots,L\}}$ are given by ${\mbfU_l\in\mathbb{C}^{M\times M_\trf}}$ and ${\mbfV_l = [\mbfv_{1l} \, \ldots \, \mbfv_{Kl}] \in\mathbb{C}^{M_\trf\times K}}$, respectively. We assume ${M_\trf= K}$ so the system can generate a different signal for each user. 
 The precoding vector ${\mbfu\sikl\in\mathbb{C}^M}$, i.e., the $k\ssth$ column of $\mbfU_l,$ is chosen from a unitary codebook with $B$ predefined beams ${\boldsymbol{\mclU}=\{\mbflu_1,\ldots,\mbflu_b,\ldots,\mbflu_B\}}$. The APs collaboratively transmit the $\mbup$ data signals $s_k$, ${\forall k\in\mathcal{K}\triangleq\{1,\ldots,K\}}$  to the users. The received signal at user $k$ during the data transmission phase is given by
\begin{equation}\label{eq:ReceivedSignalMultiBeam}
y_k=\sqrt{\frac{p_\tT}{K}}\sum_{l=1}^L\sum_{i=1}^K\mbfh\sikl^H\mbftu\siil s_k+n_k,
\end{equation}
where $\mbftu\siil\triangleq\mbfU_l\mbfv\siil$ and ${n_k\sim\mathcal{CN}(0,\sigma_n^2)}$ is the noise at user $k$, ${p_\tT>0}$ is the transmit power of an AP. Equal power allocation is assumed at the APs. The $\mbsr$ performances can be improved by another fixed but unequal power allocation solution \cite{He2017c}. Since the main goal of this paper is centered around joint design solutions with low complexity and effective beam searching and ML algorithms, the extension in this direction is omitted for the sake of simplicity.  

The received signal \eqref{eq:ReceivedSignalMultiBeam} can be rewritten as 
\begin{subequations}
\begin{align}\label{eq:ReceivedSignalRewritten}  
&&y_k=&D_k+I_k+n_k, \\ 
&\hspace{-1.5cm}\text{where}&& \nonumber\\ 
&&\!D_k\!=&\sqrt{\frac{p_\tT}{K}}\sum_{l=1}^L\mbfh\sikl^H\mbftu\sikl  s_k \text{ and }\\
&&I_k\!=&\sqrt{\frac{p_\tT}{K}}\sum_{l=1}^L\sum_{{\substack{j=1\\j\neq k}}}^K\mbfh\sikl^H\mbftu\sijl s_j
\end{align} 
\end{subequations} 
are the desired and interference signals, respectively.
\vspace{-.2cm}
\subsection{Channel Model}
The channels are characterized by using the extended $\mbSV$ geometric channel model with $P$ scatterers per user \cite{1146527} \cite{4200712}. The channel vector is defined by
\begin{equation}\label{eq:ChannelVector}
\mbfh\sikl=\sqrt{\frac{M}{P}}\sum_{p=1}^P\frac{\beta\sikl^p}{\sqrt{\alpha\sikl^p}}\mbfa(\theta\sikl^p),
\end{equation}
where ${\beta\sikl^p}$ is the channel gain with a random but fixed complex value, $\mbfa(\theta\sikl^p)$ is the array response vector for a given angle of departure (AoD), ${\theta\sikl^p\in[-\pi,\pi)}$ is the AoD for the path $p$, and ${\alpha\sikl^p}$ is the path loss given by
\begin{equation}\label{eq:PathLoss}
\alpha\sikl^p(\text{dB}) =20\log_{10}(4\pi f_c/c)+10n\log_{10}(d)+X_\sigma,
\end{equation}
where $c,f_c,n,d$, and $X_\sigma$ denote the speed of light (m/s), the carrier frequency (Hz), the path loss exponent, the distance (m), and the shadow fading following a normal distribution with mean $0$ and standard deviation $\sigma$ (dB).

Assuming each AP is equipped with ULA, the element $y,~y=1,\ldots,M$ of the array response vector for AP $l$ is given by  
\vspace{-.2cm}
\begin{equation}%\label{eq:ChannelVector}
\mbfa(\theta\sikl^p,y)=\sqrt{\frac{1}{M}}e^{(y-1)j2\pi(d/\lambda)\sin(\theta\sikl^p)}. %\right].
\end{equation}

\section{Problem Formulation}\label{sec:ProblemFormulation}
In this section, the joint design of beam selection and digital precoder design problem in $\mbcf$ networks under BCC formulated.  The beam selection metric affects the complexity of the searching algorithm.
In this section, two  beam selection metric options are introduced as well. 
\subsection{Joint Design}\label{subsec:JointDesign}
The rate of user $k$ is given as
\begin{subequations}%
\begin{align}%
\tR_k&=\log_2(1+\tSINR_k),\label{eq:Rate}\\
\text{where}\hspace{2cm}& \nonumber\\
\tSINR_k&=\frac{E\{\left|D_k\right|^2\}}{E\{\left|I_k\right|^2\}+\sigma_n^2}\nonumber\\
&=\frac{\frac{p_\tT}{K}\left|\sum_{l=1}^L\mbfh\sikl^H\mbftu\sikl\right|^2}{\frac{p_\tT}{K}\sum_{{\substack{j=1\\j\neq k}}}^K\left|\sum_{l=1}^L\mbfh\sikl^H\mbftu\sijl\right|^2+\sigma_n^2} \label{eq:SINR}
\end{align}
\label{eq:RateSINR}
\end{subequations}
is the signal-to-interference-plus-noise-ratio (SINR) of user $k$. 

The joint design problem of beam selection and digital precoder under BCC for $\mbsr$ maximization is given as
\begin{subequations}\label{op:SumRateMultiBeam}
\begin{align}
&~\underset{\{\mbfu\sikl\}}{\arg\max}~\underset{\{\mbfv\sikl\}}{\max}&&\sum_{k=1}^K \tR_k \label{op:SumRateMultiBeama} \\ 
&~~\tst&&\hspace{-1.5cm} \mbfu\sikl\!\in\boldsymbol{\mclU},\forall k\in\mathcal{K},\forall l\in\mathcal{L}, \label{op:SumRateMultiBeamb} \\ 
&&&\hspace{-1.5cm}\mbfu\sikl\neq\mbfu\sijm,\forall k,\forall j, j\!\neq\!k\!\in\mathcal{K},\forall l,\forall m\!\in\mathcal{L}, \label{op:SumRateMultiBeamc} \\ 
&&&\hspace{-1.5cm}\|\mbfU_l\mbfv\sikl\|_2^2=1,\forall k\in\mathcal{K},\forall l\in\mathcal{L}. \label{op:SumRateMultiBeamd}
\end{align}
\end{subequations}
The first constraint \eqref{op:SumRateMultiBeamb} implies that the analog precoders are chosen from a given codebook. The second constraint \eqref{op:SumRateMultiBeamc} asserts BCC on the solution, i.e., the same beam cannot be assigned to multiple users. Note that in $\mbcf$ networks, all APs collaboratively transmit the same data signal to a user. Therefore, as seen in \eqref{op:SumRateMultiBeamc}, APs are allowed to choose the same beam to the same user. Finally, the last constraint \eqref{op:SumRateMultiBeamd} asserts that the digital precoder does not provide a power gain.   

\subsection{Selection Metrics}\label{subsec:SelectionMetrics}
A beam selection algorithm that is based on the maximum received signal power at each user can be achieved distributively with a low computational complexity and with low signalling overhead. The selection metric as the received signal power between AP $l$ and user $k$ is given by
\begin{equation}\label{eq:LinkPower}
E\{\left|D\sikl\right|^2\}=\frac{p_\tT}{K}\left|\mbfh\sikl^H\mbftu\sikl\right|^2.
\end{equation}

However, the $\mbsr$ reflects the $\mbto$ between strong signal and weak interference which is the essence of $\mbsr$ maximization problem given in \eqref{op:SumRateMultiBeam} \cite{Amadori2015a,Ren2017}. ML algorithms provide the needed succor for the increased costs due to using the $\mbsr$ as the beam selection metric. In other words, since the training of ML algorithms is performed $\mbol$, the increased costs can be tolerated. 

\section{Proposed Joint Design Algorithms}\label{sec:BeamSelection}
The beam selection problem is distinctively challenging. In a $\mbcf$ network, there is only one large cell where there are far more APs than users, and all APs serve all users. An exhaustive algorithm centrally selects the optimal beams by evaluating the $\mbsrs$ achieved by all stream combinations between all APs and users, and then by choosing the beam combination that results in the maximum $\mbsr$. The complexity of an exhaustive algorithm is $B^{KL}$.  
Assuming the time slot of beam transmission is normalized to $1$, the delay of a beam selection algorithm is equal to its complexity. For instance, an exhaustive algorithm awaits $B^{KL}$ time slots to evaluate all beam combinations and centrally select the best beam combination. Therefore, as the $\mbcf$ network size grows, an exhaustive algorithm becomes intractable in its computational complexity and initial access delay. 

\begin{figure}[!t]
\centering
 \begin{overpic}[height=6.5cm, width=5.75cm] {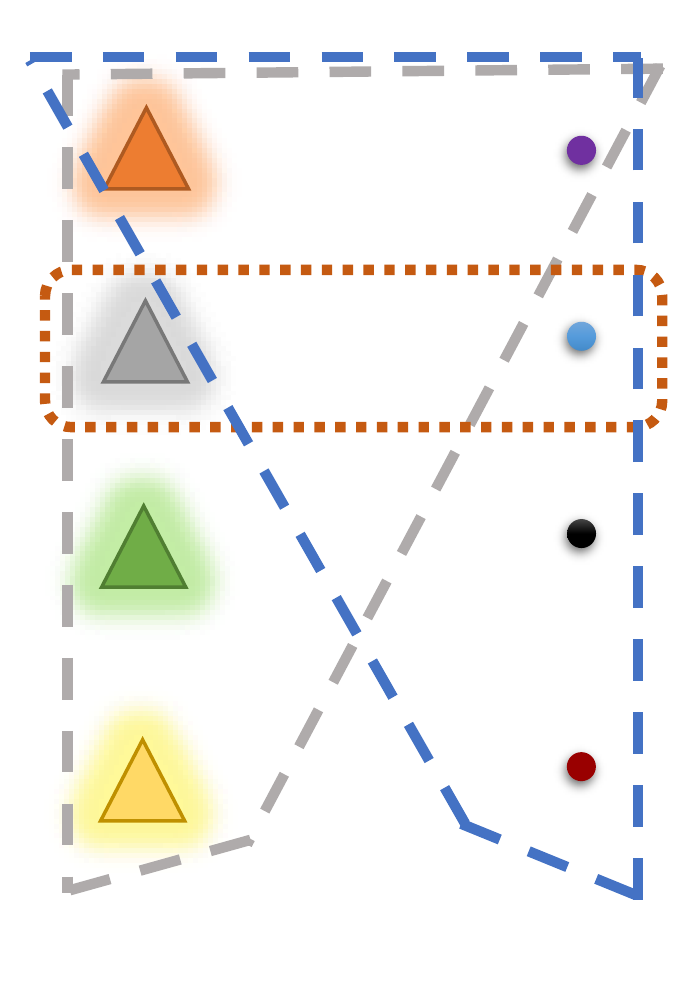}
\put(15,5){\footnotesize\textbf{APs}} \put(68,5){\footnotesize\textbf{Users}}
\put(10,34){\color{gray}\footnotesize\textbf{Semicentralized}}
\put(12,30){\color{gray}\footnotesize\textbf{search segment}}
\put(55,34){\color{MyBlue}\footnotesize\textbf{Semilinear}}
\put(52,30){\color{MyBlue}\footnotesize\textbf{search segment}}
\put(38,66){\color{brown}\footnotesize \textbf{Linear}}
\put(31,62){\color{brown}\footnotesize \textbf{search segment}}
 \end{overpic}
\caption{The search segments of $\mbsc$, $\mbslin$ and linear search algorithms. The search segment of centralized, i.e., exhaustive, algorithm covers the whole network.}
\label{Fig:SearchSegments}
 \end{figure}
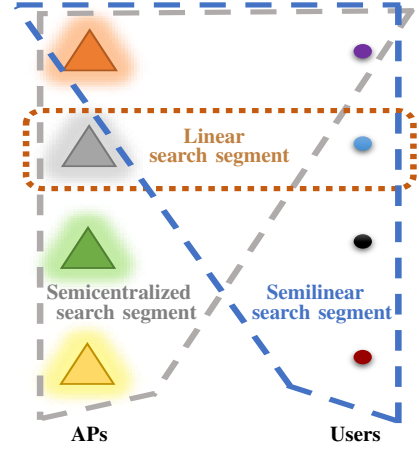

In this section, we propose two low complexity, thus low delay, beam selection algorithms, which are coined as $\mbslin$ and linear search algorithms. The search processes are iterated segment by segment as illustrated in Fig. \ref{Fig:SearchSegments}. While  all possible beam combinations are  being searched in a segment, beams in other segments are held fixed. In the $\mbslin$ algorithm, a segment is formed between an AP and all users. In other words, compared to the centralized algorithm where the search segment is the whole network, the searching process is segmented into APs. To reduce the complexity further, in the linear search algorithm, the searching process is segmented into APs as well. Hence, the complexities of $\mbslin$ and linear search algorithms become linear in only $L$, and in both $L$ and $K$, respectively. 

\subsection{Semilinear and Semicentralized Searches}\label{subsec:SemiLinearandSemiCentralizedSearches}
 In the $\mbslin$ search algorithm, the selection process is proceeded between each AP and all users. After the transmissions of $B^K$ possible beam combinations from an AP to all users, the beam combination that achieves the maximum $\mbsr$ is selected. Then, the algorithm proceeds with the selection of the best beam combination from the next AP to all users. The selection of beams from an AP to all users is carried out by fixing the beams from all other APs $\forall l^\prime,~l^\prime \in\{1,\ldots,L\}\backslash\{l\}$ in the network. The complexity of this algorithm is $L(B^K)$, which is linear in the number of APs. 

On the other hand, in the $\mbsc$ search algorithm, the selection process is proceeded between all APs and a user. After the transmissions of $B^L$ possible beam combinations from all APs to user $k$, the beam combination that achieves the maximum  rate at user $k$ is selected. Then, the algorithm proceeds with the selection of the best beam combination for the next user. The selection  of beams from all APs to user $k$ is carried out by fixing the beams to other users $\forall k^\prime,~k^\prime \in\{1,\ldots,K\}\backslash\{k\}$ in the network. The complexity of this algorithm is $K(B^L)$, which is linear in the number of users.
  
For the joint design problem \eqref{op:SumRateMultiBeam}, $\mbslin$ is advantageous over the $\mbsc$ search algorithm from two aspects. First of all, as mentioned earlier, in $\mbcf$ networks, ${L\gg K}$. Hence, a search algorithm which is linear in $L$ is preferable. Secondly, when ${B<L}$, the $\mbsc$ algorithm can run out of available beams for other users $k^\prime$ after the first search segment in case each AP wishes to transmit a different beam to user $k$. Note that, due to BCC, for both $\mbslin$ and $\mbsc$ algorithms ${B\geq K}$ must be satisfied.

Next, a linear search algorithm is proposed to achieve a more fundamental complexity reduction. 

\subsection{Linear Search}\label{subsec:LinearSearch}
In the linear search algorithm, a pair of one AP and one user forms a search segment. The selection process is proceeded pair by pair. 
The selection of beam from AP $l$ to user $k$ is carried out by fixing all other the beams in the network. In particular, the beams between AP $l$ to users $\forall k^\prime$, and the beams between APs $\forall l^\prime,~l^\prime\in\{1,\ldots,L\}\backslash\{l\}$ and users $\forall k$ are fixed. The complexity of this algorithm is $LKB$, which is linear in the number of APs, users, and antennas at each AP.

\subsection{Beam Conflict Control}\label{subsec:BCC}
Searching processes at the segments of the proposed searching algorithms can be executed in parallel or series. The latter option improves the search quality, but it introduces a delay. However, due to BCC, parallel search can introduce even more delays and it can make the search process even more complex as follows. The beam selection results of the parallel search segments need to be immediately announced in the network. The segments that receive the announcements before the start of the search process can update the codebook log for each user $\boldsymbol{\mathcal{U}}_k$, i.e., remove a codeword from the codebook of user $k$, $\boldsymbol{\mathcal{U}}_k$, if that codeword is assigned to another user $k^\prime$. If the searching process has already started, then the result of the process can be waited. If the result leads a codeword assignment that is already assigned to another user, then the search process needs to be restarted after updating $\boldsymbol{\mathcal{U}}_k$. Hence, whether synchronous or asynchronous parallel search process is applied, the delay of parallel process can be even more than the serial process under BCC.

Note that BCC is different in $\mbcf$ networks than in other networks where the same user is not coherently served by the APs, i.e., APs transmit different data signals to the same user. In other networks, between each AP and a user, a different analog beam should be used. In this case, once a beam is selected, it can be removed from the common codebook $\boldsymbol{\mathcal{U}}$. In $\mbcf$ networks, all APs want to coherently transmit to each user. Therefore, between all APs and a user, the same analog beam can be used if APs choose to, but the same beam cannot be used for another user. In this case, once a beam is selected, it should not be removed from the codebook in case another AP wants to use it for the same user. This means that  a codebook log for each user $\boldsymbol{\mathcal{U}}_k$ must be updated and announced after each decision in the network. This makes BCC a more challenging task in $\mbcf$ networks.

However, under BCC, the number of beam combinations is reduced from ${LB^K}$ to ${LB!/(B-K)!}$ for the linear search algorithm. Hence, without BCC, it is more costly to try all combinations as $B$ and $K$ increase. As demonstrated numerically in Section \ref{sec:NumericalResults}, the algorithms with and without BCC, and the algorithms with BCC initialization result in interesting $\mbox{trade-offs}$ between the network $\mbsr$ and simulation duration.

\subsection{Digital Precoders}\label{subsec:DigitalPrecoder}
In this work, we utilize ZF digital precoders in the baseband process of APs. At AP $l$, ZF precoder is given as
\begin{subequations}\label{eq:ZFFilter}
\begin{align}
&\mbfV_l^\text{ZF}=\mbfH^H_l\left(\mbfH_l\mbfH_l^H\right)^{-1},\\
&\hspace{-2.8cm}\text{where} \nonumber\\
&\mbfH_l=\begin{bmatrix} \mbfh_{1l}^H\mbfU_l \label{eq:EffectiveChannel} \\
\cdots \\
\mbfh_{Kl}^H\mbfU_l \\
\end{bmatrix}
\end{align}
\end{subequations}
is the effective channel from AP $l$ to all users. 

The MMSE precoder can be preferred in the lower signal-to-noise ratio (SNR) regime where it can achieve higher $\mbsr$ results than the ZF precoder.
Furthermore, when BCC is not used, $\mbfH_l$ can be a $\mblr$ matrix. Thus, MMSE precoder can be used instead of ZF precoder since it can support the $\mblr$ issue to some extent. However, BCC is worthwhile since the $\mbsr$ results without BCC are lower as demonstrated by numerical results in Section \ref{sec:NumericalResults}. The MMSE precoder at AP $l$ is given as
\begin{equation}\label{eq:MMSEFilter}
\mbfV_l^\text{MMSE}=\mbfH^H_l\left(\frac{p_\tT}{K}\mbfH_l\mbfH_l^H+\sigma_n^2\mbfI_K\right)^{-1}, 
\end{equation}
where $\mbfI_K$ is the identity matrix of size ${K\times K}$.

The digital precoders in \eqref{eq:ZFFilter} and \eqref{eq:MMSEFilter} require local baseband CSI, i.e., the effective channel vectors from AP $l$ to all users, ${\mbfh\sikl^\text{eff}=\mbfh\sikl^H\mbfU_l}$, ${\forall k\in\mathcal{K}}$. Here, ${(\mbfh\sikl^\text{eff})^T\in\mathbb{C}^{M_\trf}}$,  where $(.)^T$ is the transpose operator. Note that the channel ${\mbfh_{kl}\in\mathbb{C}^M}$ cannot be directly estimated due to the constraint in the number of RF chains, ${M_\trf\leq M}$, in general \cite{Hu2018}. Different analog beamforming vectors result in different effective channels. Therefore, as widely proposed in the literature and standards, after the beam selections are concluded, the effective channels are estimated to design the digital precoders in the final stages of the hybrid designs \cite{8322248,8630933,Han2018,8599167,Amadori2015a}. We refer this approach as a disjoint design. On the other hand, as detailed in the next section, our proposed joint designs require the effective channel estimation and digital precoder design at AP $l$ each time AP $l$ tests a beam. Similar to multiple initializations, iterations, and selection features, the mentioned requirement can be impractical in fast varying channel conditions. However, as detailed in Section \ref{sec:SupervisedML} and numerically demonstrated in Section \ref{sec:NumericalResults}, ML algorithms can overcome these impractical challenges. By training the ML algorithms $\mbol$ with our proposed $\mbwc$ designs to do the beam selection, online execution of the ML algorithms for the beam selection followed by a digital precoder design mimics the proposed $\mbwc$ designs and can achieve nearly the same $\mbsr$ results at significantly lower computational complexities.

\begin{algorithm}[t]\small
\caption{Pseudocodes of the proposed semilinear-II-rate and linear-II-rate algorithms.}\label{alg:SearchAlgs}
\begin{algorithmic}[1]
\State $\tR_\ts^\tm=0$
\For  {$i=1:\mathrm{Initializations}$} 
\State Random initializations of analog precoders ${\mbfu\sikl^0\in{\boldsymbol{\mclU}}}$, ${\forall k,\forall l}$ \label{step:SearchAlgsInit}
\Statex \hspace{.5cm}$\mbfu\sikl^0\neq \mbfu_{k^\prime j}^0, \forall l,\forall j, \boldsymbol \forall k,\forall k^\prime, k\neq k^\prime$ (with BCC)
\State $\mbfu\sikl=\mbfu\sikl^0$, $\forall k, \forall l$ 
\State $\mbflU^\tcomb$ (for the semilinear search algorithm) \label{step:CodebookSemiLin}
\State $\boldsymbol{\mathcal{U}}_k=\boldsymbol{\mclU}$, $\forall k$ (for the linear search algorithm) \label{step:CodebookAlg1}
\For {$s=1:\mathrm{Iterations}$} \label{step:Iterations}
\For {$l=1:L$} 
\State \hspace{-.4cm} \textbf{\textit{Choose semilinear or linear search algorithm}} \label{step:Option}
\EndFor %l
\State $\tR_\ts^\prime$  \label{alg:SameLinesA}
\If {$\tR_\ts^\prime>\tR_\ts^{\tm}$}
\State $\mbfU_l^\topt\!=\!\mbfU_l^\star$, $\forall l$ (for the semilinear search algorithm)
\State $\mbfu\sikl^{\topt}=\mbfu\sikl^\star$, $\forall k,\forall l$ (for the linear search algorithm)
\State $\tR_\ts^{\tm}=\tR_\ts^\prime$
\EndIf   \label{alg:SameLinesB}
\EndFor %s
\EndFor %i
\end{algorithmic}
\end{algorithm}

\begin{algorithm}[t]\small
\caption{The proposed $\mbslin$ search algorithm.}\label{alg:SearchAlgsA}
\begin{algorithmic}[1]
\For {$c=1:C$} \label{step:SemiLinearStart}
\State $\mbfU_l=\mbflU_c^\tcomb$ \label{step:SemiLinearJointa}
\State $\mbfV_l$ \label{step:SemiLinearJointb}
\State $\tR_\ts(c)$
\EndFor %o
\State $[\sim , c^\star]=\text{maxvec}\left[\tR_\ts(1)\ldots\tR_\ts(c)\ldots\tR_\ts(C)\right]$  \label{step:maxvector}
\State $\mbfU_l^\star=\mbflU_{c^\star}^\tcomb$ 
\State Update $\mbflU^\tcomb$ (with BCC) \label{step:SemiLinearEnd}
\State Update $C$ (with BCC) \label{step:SemiLinearCUpdate}
\end{algorithmic}
\end{algorithm}
  
\begin{algorithm}[t]\small
\caption{The proposed linear search algorithm.}\label{alg:SearchAlgsB}
\begin{algorithmic}[1]
\For {$k=1:K$} \label{step:LinearStart}
\For {$b=1:B_k$} \label{step:LinearStartM}
\State $\mbfu\sikl=\mbflu_b$  \label{step:LinearJointa}
\State $\mbfV_l$  \label{step:LinearJointb}
\State $\tR_\ts(b)$  
\EndFor % m
\State $[\sim , b^\star]=\text{maxvec}\left[\tR_\ts(1)\ldots\tR_\ts(b)\ldots\tR_\ts(B_{k})\right]$  
\State $\mbfu\sikl^\star=\mbflu_{b^\star}$ 
\State ${\boldsymbol{\mathcal{U}}_{k^\prime}\leftarrow{\boldsymbol{\mathcal{U}}_{k^\prime}\backslash~\mbflu_{b^\star}}}$~, $\forall k^\prime,k^\prime\neq k$ (with BCC)\label{step:BCCLinear}
\State Update $B_{k^\prime}$, $\forall k^\prime,k^\prime\neq k$ (with BCC) \label{step:UpdateM} 
\EndFor \label{step:LinearEnd} %k  
\end{algorithmic}
\end{algorithm}

\begin{algorithm}[t]\small
\caption{Pseudocode of the proposed linear-IIS-rate algorithm.}\label{alg:SearchAlgswS}
\begin{algorithmic}[1]
\State $\tR_\ts^\tm=0$
\For  {$i=1:\mathrm{Initializations}$} \label{step:RandomInitializations}
\State Random initializations of analog precoders ${\mbfu\sikl^0\in{\boldsymbol{\mclU}}}$, ${\forall k,\forall l}$ \label{step:RandInit}
\Statex \hspace{.5cm}$\mbfu\sikl^0\neq \mbfu_{k^\prime j}^0, \forall l,\forall j, \boldsymbol \forall k,\forall k^\prime, k\neq k^\prime$ (with BCC)
\State $\boldsymbol{\mathcal{U}}_k=\boldsymbol{\mclU}$, $\forall k$ \label{step:CodebookAlg4}
\State $\mbfL\sivec=1:L$ \label{step:LvecInit}
\For {$j=1:L$}  \label{step:PrioritizedSegment} 
\State $\mbfu\sikl=\mbfu\sikl^0$, $\forall k, \forall l$ \label{step:AnalogPrecoderReset}
\State  $\tR_\ts^{\tm^\prime}=0$
\For {$s=1:\mathrm{Iterations}$} \label{step:IterationswP} 
\For {$l\in \mbfL\sivec$} 
\State \hspace{-.4cm} \textbf{\textit{Apply linear search algorithm}} \label{step:OptionwS} 
\EndFor %l
\State Same lines \ref{alg:SameLinesA}-\ref{alg:SameLinesB} in Algorithm \ref{alg:SearchAlgs}
\EndFor 
\If {$\tR_\ts^{\tm^\prime}>\tR_\ts^\tm$} 
\State $\mbfu\sikl^\topt=\mbfu\sikl^{\topt^\prime}$, $\forall k,\forall l$ 
\State $\tR_\ts^\tm=\tR_\ts^{\tm^\prime}$ 
\EndIf 
\State $\mbfL\sivec=\mathrm{circshift}(\mbfL\sivec)$ \label{step:CircularShift}
\EndFor %j
\EndFor %i
\end{algorithmic}
\end{algorithm}

\subsection{Pseudocodes}\label{subsec:Pseudocode}
The pseudocodes of the proposed $\mbslin$ and linear search algorithms are given in Algorithms \ref{alg:SearchAlgs}-\ref{alg:SearchAlgsB}. The common lines of the proposed algorithms are given in Algorithm \ref{alg:SearchAlgs}. At step \ref{step:Option} of Algorithm \ref{alg:SearchAlgs}, either the $\mbslin$ or linear search in Algorithm \ref{alg:SearchAlgsA} or \ref{alg:SearchAlgsB} is chosen, respectively. In Algorithm \ref{alg:SearchAlgswS}, the pseudocode of the proposed algorithm for the prioritization of search segments is given. For brevity, based on the proposed joint design of beam selection and digital precoder, the algorithms that apply multiple initializations and iterations are denoted by II in Algorithm \ref{alg:SearchAlgs} and the algorithm that applies the selection method as well is denoted by IIS in Algorithm \ref{alg:SearchAlgswS}. Algorithms that use the DL power in \eqref{eq:LinkPower} and the $\mbsr$ as the beam selection metrics are noted by DL and rate, respectively. The pseudocodes of algorithms that use DL metric are same as in Algorithms \ref{alg:SearchAlgs}-\ref{alg:SearchAlgswS} after replacing the $\mbsr$ metrics, ${\tR_\ts\triangleq\sum_{k=1}^K\tR_k}$ with the \mbox{sum-DL} metrics, ${\tDL_\ts\triangleq\sum_{l=1}^L\sum_{k=1}^KE\{\left|D\sikl\right|^2\}}$.
 
The proposed algorithms are $\mbwc$ to train ML algorithms in the sense that they are consolidated from joint design of analog and digital beam precoders (lines \ref{step:SemiLinearJointa}-\ref{step:SemiLinearJointb} for the $\mbslin$ search in Algorithm \ref{alg:SearchAlgsA} and lines \ref{step:LinearJointa}-\ref{step:LinearJointb} for the linear search in Algorithm \ref{alg:SearchAlgsB}), multiple random initializations of analog precoders (the \mbox{for-loops} at step \ref{step:RandomInitializations} in Algorithms \ref{alg:SearchAlgs} and \ref{alg:SearchAlgswS}), multiple iterations of the joint design (the \mbox{for-loops} at steps \ref{step:Iterations} and \ref{step:IterationswP} in Algorithms \ref{alg:SearchAlgs} and \ref{alg:SearchAlgswS}, respectively), and finally, the selection of the prioritized search segment to assign the beams first than the other segments (the $\mbfl$ at step \ref{step:PrioritizedSegment} in Algorithm \ref{alg:SearchAlgswS}).

As explained earlier, a codebook log is updated and announced at each user for BCC implementation in the $\mbcf$ network. As seen at steps \ref{step:CodebookAlg1} and \ref{step:CodebookAlg4} of the Algorithms \ref{alg:SearchAlgs} and \ref{alg:SearchAlgswS}, respectively, the codebook log at each user $\boldsymbol{\mathcal{U}}_k$ is initialized by copying the codebook $\boldsymbol{\mclU}$. On the other hand, for the $\mbslin$ search algorithm, a single codebook log, ${\mbflU^\tcomb}$, is tracked for simplicity as seen at step \ref{step:CodebookSemiLin} of the Algorithm \ref{alg:SearchAlgs}. ${\mbflU^\tcomb}$ is the matrix set of beam combinations under BCC
\begin{equation}\label{eq:Ucomb1} 
\mbflU^\tcomb\triangleq\left\{\mbflU^\tcomb_1,\ldots,\mbflU^\tcomb_c,\ldots,\mbflU^\tcomb_C\right\},
\end{equation}
where initially, ${C=B(B-1)\ldots(B-K+1)}$ and ${\mbflU^\tcomb_c\in\mathbb{C}^{M\times K}}$ is given as 
\begin{equation}\label{eq:Ucomb2} 
\mbflU^\tcomb_c=\left[\mbfu_{i_c(1)}\ldots\mbfu_{i_c(K)}\right].
\end{equation}
Here, $i_c(k)$ is the $c\ssth$ beam index combination for user $k$ such that ${i_c(k)\neq i_c(k^\prime),\forall k,\forall k^\prime, k\neq k^\prime}$. 

In Algorithm \ref{alg:SearchAlgsA}, the $\mbfl$ at step \ref{step:SemiLinearStart} of the $\mbslin$ search algorithm tests the combinations from AP $l$ to all users. At step \ref{step:SemiLinearJointa}, $\mbflU_c^\tcomb$ is the $c\ssth$ combination for the precoding matrix with the available codeword combinations at the columns to be tested as given in the equations \eqref{eq:Ucomb1} and \eqref{eq:Ucomb2}. At step the \ref{step:maxvector}, maxvec operator returns the index of the maximum vector element. When the index combination that yields the maximum $\mbsr$ is obtained at step \ref{step:maxvector}, i.e., $\mbflU^\tcomb_{c^\star}$, ${\mbflU^\tcomb}$ is updated at step \ref{step:SemiLinearEnd} as follows
\begin{equation}
\mbflU^\tcomb\leftarrow \mbflU^\tcomb~\backslash~\mbflU^\tcomb_c,~\forall c \mid i_{c^\star}(k)= i_c(k^\prime),\forall k,\forall k^\prime, k\neq k^\prime.
\end{equation}

Assume ${K=2}$, ${B=3}$, and no beam assignment is realized yet. Then the total number of beam combinations to be tested under BCC is $6$, i.e., ${C=6}$. Hence, the matrix set of beam combinations is initially given as $${\mbflU^\tcomb=\{[\mbflu_1 \mbflu_2],[\mbflu_1 \mbflu_3],[\mbflu_2 \mbflu_1],[\mbflu_2 \mbflu_3],[\mbflu_3 \mbflu_1],[\mbflu_3 \mbflu_2]\}}.$$ Assume ${c^\star=4}$, hence ${\mbflU_{c^\star}^\tcomb=[\mbflu_2 \mbflu_3]}$. At step \ref{step:SemiLinearEnd}, the matrix set of beam combinations is updated as follows $${\mbflU^\tcomb=\{[\mbflu_1 \mbflu_3],[\mbflu_2 \mbflu_1],[\mbflu_2 \mbflu_3]\}}.$$ Finally, at step \ref{step:SemiLinearCUpdate}, ${C=3}$ update is performed. 

For the linear search algorithm, similar steps are executed except that in a search segment, the beam combinations between an AP and a user are tested each time as seen in the \mbox{for-loops} at steps \ref{step:LinearStart} and \ref{step:LinearStartM} of the Algorithm \ref{alg:SearchAlgsB}. At step \ref{step:LinearStartM} of Algorithm \ref{alg:SearchAlgsB}, $B_k, \forall k\in\mathcal{K}$, are initially equal to $B$. However, due to BCC, as codebooks are removed from users' codebook logs, the number of possible beam selections for user $k$ is updated at step \ref{step:UpdateM}. At step \ref{step:BCCLinear}, the selected codeword $\{\mbfu\sikl^\star\}$ is removed from the codebook logs of all others users $k^\prime$, ${\boldsymbol{\mathcal{U}}_{k^\prime}}$, for BCC. Hence, when the codeword combinations for the other users $k^\prime$ are being tested next time, the already assigned beam for user $k$ does not exist in ${\boldsymbol{\mathcal{U}}_{k^\prime}}$, ${\forall k^\prime, k^\prime\neq k}$ as an option to be tested. Although the BCC implementations are achieved in a single line at steps \ref{step:SemiLinearEnd} and \ref{step:BCCLinear} of the Algorithms \ref{alg:SearchAlgsA} and \ref{alg:SearchAlgsB}, respectively, their coding implementations are significantly challenging \cite{YetisGitHubJointDesign}.

At step \ref{step:AnalogPrecoderReset} of Algorithm \ref{alg:SearchAlgswS}, the analog precoders are reset back to the random initializations that were determined at step \ref{step:RandInit}. The order of searching process is determined by the $\mbfL\sivec$ vector that is initialized at step \ref{step:LvecInit} and updated at step \ref{step:CircularShift} with a circular shift. For instance, during the first iteration of the $\mbfl$ at step \ref{step:PrioritizedSegment}, the search segment-1, i.e., AP $1$, does the assignment first since $\mbfL\sivec=[1,2,\ldots,L]$. Whereas, during the second iteration of the $\mbfl$, the search segment-2, i.e., AP $2$, does the assignment first, since $\mbfL\sivec=[2,\ldots,L,1]$. When the loop at step \ref{step:PrioritizedSegment} is finalized, the priority is given to the search segment $l$, i.e., AP $l$, with the highest $\mbsr$. To improve the $\mbsr$  further at an increased cost, each segment can have a priority order. For simplicity, in this work, only one search segment is prioritized, e.g., if the search segment-2 is prioritized then the search segments $3,\ldots,L,$ and, finally, $1$ are executed in the written order. 

As supported by the extensive numerical analyses in Section \ref{sec:NumericalResults}, the $\mbsrs$ and complexities of the proposed algorithms are given as follows: linear-IIS-rate (Algorithm \ref{alg:SearchAlgswS}) $>$ semilinear-II-rate (Algorithms \ref{alg:SearchAlgs} and \ref{alg:SearchAlgsA}) $>$ linear-II-rate (Algorithms \ref{alg:SearchAlgs} and \ref{alg:SearchAlgsB}) $>$ linear-II-DL (Algorithms \ref{alg:SearchAlgs} and \ref{alg:SearchAlgsB} with DL metrics instead). 

At an increased cost, a $\mbslin$ search option can be added to the step \ref{step:OptionwS} in Algorithm \ref{alg:SearchAlgsB} as well, i.e., a semilinear-IIS-rate algorithm. It can be clearly interpolated from the numerical results in Section \ref{sec:NumericalResults} that the semilinear-IIS-rate can achieve the highest $\mbsr$ with an increased simulation duration. The selection feature can be more advantageous in asymmetric networks, i.e., the channels between APs and users are not uniformly distributed, than in symmetric networks. However, the user fairness needs to be carefully addressed \cite{4531580}.

To the best of our knowledge, in the literature and standards including IEEE 802.11ad and IEEE 802.15.3c, the beam training solutions do not consider the joint design of analog beam selection and digital precoder. Hence, our proposed solutions are benchmarked with the disjoint design given in Algorithm \ref{alg:SearchAlgsDisjoint}. As seen at step \ref{step:Disjointb}, the digital precoders are designed after the selections of analog beams are finalized. In contrast, in our proposed solutions, the digital precoders are designed at each iteration of the analog beam search process as seen at steps \ref{step:SemiLinearJointb} and \ref{step:LinearJointb} of the Algorithms \ref{alg:SearchAlgsA} and \ref{alg:SearchAlgsB}, respectively. 

Finally, the algorithms without BCC implementation can be obtained by inactivating the steps \ref{step:SemiLinearEnd} and \ref{step:BCCLinear} in Algorithms \ref{alg:SearchAlgsA} and \ref{alg:SearchAlgsB}, respectively, and also by inactivating the condition that the random initializations must satisfy BCC in the initialization steps \ref{step:SearchAlgsInit}, \ref{step:RandInit}, and \ref{step:SearchAlgsDisjointInit} of the Algorithms \ref{alg:SearchAlgs}, \ref{alg:SearchAlgswS}, and \ref{alg:SearchAlgsDisjoint}, respectively.  

\begin{algorithm}[t]\small
\caption{Pseudocode of the benchmark disjoint linear-DL algorithm.}\label{alg:SearchAlgsDisjoint}
\begin{algorithmic}[1]
\State Random initializations of analog precoders ${\mbfu\sikl^0\in{\boldsymbol{\mclU}}}$, ${\forall k,\forall l}$ \label{step:SearchAlgsDisjointInit}
\Statex $\mbfu\sikl^0\neq \mbfu_{k^\prime j}^0, \forall l,\forall j, \boldsymbol \forall k,\forall k^\prime, k\neq k^\prime$ (with BCC)
\State $\boldsymbol{\mathcal{U}}_k=\boldsymbol{\mclU}$, $\forall k$ 
\State $\mbfu\sikl=\mbfu\sikl^0$, $\forall k, \forall l$ 
\For {$l=1:L$} 
\For {$k=1:K$} 
\For {$b=1:B_k$} 
\State $\mbfu\sikl=\mbflu_b$  
\State $\tDL_\ts(b)$  
\EndFor % m
\State $[\sim , b^\star]=\text{maxvec}\left[\tDL_\ts(1)\ldots\tDL_\ts(b)\ldots\tDL_\ts(B_{k})\right]$  
\State $\mbfu\sikl^\star=\mbflu_{b^\star}$ 
\State ${\boldsymbol{\mathcal{U}}_{k^\prime}\leftarrow{\boldsymbol{\mathcal{U}}_{k^\prime}\backslash~\mbflu_{b^\star}}}$~, $\forall k^\prime,k^\prime\neq k$ (with BCC)
\State Update $B_{k^\prime}$, $\forall k^\prime,k^\prime\neq k$ (with BCC) 

\EndFor %k  
\EndFor %l
\State $\mbfV_l$, $\forall l$ \label{step:Disjointb}
\end{algorithmic}
\end{algorithm}

\section{Supervised Machine Learning Algorithms}\label{sec:SupervisedML}
Among supervised ML algorithms, SVM, MLP, and RFt are prevalent algorithms. These algorithms are powerful, flexible, and swift with regards to their good accuracies, many hyperparameters, and light computational complexities, respectively.

In this section, we discuss $\mbmo$ classification and the effect of BCC on it, and provide Scikit implementation details.

\subsection{Multi-Output Classification}\label{sec:MOMC}
Single-output classification methods select one class at a time. On the other hand, $\mbmo$ classification methods select multiple classes at once. In the literature, the latter method is also  referred as $\mbml$ $\mbmc$ classification method, or shortly, $\mbml$ classification method. 

Between an AP and all users, there are $B^K$ beam options, i.e., the number of classes is $B^K$. The final decision vector at AP $l$ in $\text{base-}B$ notation can be given as 
\begin{equation}\label{eq:labels}
\mbfb_l=\left[b_{1l} b_{2l}\ldots b\sikl \ldots b_{K-1l} b_{Kl} \right],
\end{equation}
where ${b\siil\in\{0,1,\ldots,B-1\}}$. For instance, consider a network with ${B=3}$ and ${K=2}$. There are ${B^K=9}$ classes. In $\text{base-}2$ notation, these classes are labeled as $00, 01,02,\ldots,22$. Hence, beam selection becomes a \mbox{${K=2}$-label} \mbox{${B=3}$-class} classification problem. 

The naive approach to solve $\mbmo$ classification problems is to build a decision model for each output. Hence, the problem is divided into $K$ $\mbso$ classification problems.  In the literature, the naive approach of treating each output independently from other outputs and solving a classification problem for each output is known as binary relevance. The disadvantage of this approach is that the correlations between the outputs, i.e., users, are neglected.

To accommodate the correlation between the labels, classifiers can be chained. In this approach, the first classifier is trained by the training data only and the label decisions of the first classifier are noted. Then, for the second classifier, the training data and the label decision of the first classifier are used as inputs. This is iterated until the last classifier, i.e., the last output is trained based on the training data and also, the label decisions of the previous classifiers. Clearly, classifier chains can exploit the correlation between the labels to some extent.

As outlined in \cite{101007,Tsoumakas2007}, binary relevance, classifier chains, and label \mbox{power-set} are known as problem transformation methods. Problem transformation methods transform the $\mbml$ problem into several $\mbsl$ problems and apply a $\mbsl$ algorithm. On the other hand, algorithm adaptation methods, such as $\mbml$ \mbox{$k$-nearest} neighbor \mbox{(ML-kNN)}, AdaBoost, and RFt, extend the $\mbsl$ solutions to $\mbml$ solutions. Random forests can perform $\mbml$ classification \cite{Liu2015}. In other words, it can classify all outputs at once. At each leaf, all outputs are stored rather than only a single output. Then, the impurity metric, e.g., Gini, at a split is evaluated by averaging the impurities of the outputs at that split \cite[1.10.3 \mbox{Multi-output} problems]{scikit}. Since the splits are optimized by considering all outputs, RFt can exploit the correlation between the outputs to some extent as well. In addition, a single model is generated by considering all labels. This significantly reduces the computational complexity compared to the naive approach of generating a model for each of the outputs as mentioned earlier.  
 
Due to BCC, the decision vectors $\mbfb_l$ \eqref{eq:labels} at the APs become correlated. In other words, the classification outputs of the solution to the problem \eqref{op:SumRateMultiBeam} follows a pattern. For instance, when ${B=K}$, the decision vectors at APs are equal, ${\mbfb_l=\mbfb_{l^\prime}}$, $\forall l$, $\forall l^\prime$, ${l\neq l^\prime}$. In this case, the overall decision output follows a fully repetitive pattern 
\begin{equation}\label{eq:BequalsK}
\mbfb=\left[\mbfb_1\ldots \mbfb_L\right]=\left[b_1\ldots b_k \ldots b_K\ldots b_1\ldots b_k \ldots b_K\right],
\end{equation}
where the subindex $l$ is omitted since the decisions at user $k$ are equal at all APs.

As illustrated by extensive numerical results in Section \ref{sec:NumericalResults}, RFt with the classifier chains can effectively exploit the correlated outputs occurred due to BCC. Since the outputs follow a full pattern when ${B=K}$, RFt with the classifier chains can achieve the original, i.e., $100\%$, $\mbsr$ that is achieved by the beam selection algorithms. When ${B>K}$, it is observed that RFt with the classifier chains can achieve $99\%$ of the original $\mbsr$ in contrast to $63\%$ when the classifier chains are not utilized with RFt.

BCC has a significant affect on the minimum number of beam search options depending on the network architecture. As seen in \eqref{eq:BequalsK}, the minimum number of beams in a $\mbcf$ network, $B_\tmin$, is $K$. In contrast, for an interference network, ${B_\tmin=LK}$. Hence, BCC can be beneficial in $\mbcf$ networks to reduce the beam search time. Moreover, if a DFT codebook is used, where ${M=B}$, the minimum number of transmit antennas is also significantly low.

Finally, as mentioned earlier, BCC creates highly correlated outputs in $\mbcf$ networks compared to the other networks such as interference networks. Assume a network with $L=2$, $K=4$. For $\mbcf$ and interference networks, $B_\tmin=4$ and $B_\tmin=8$, respectively. Then, the overall decision outputs for $\mbcf$ and interference networks are given as
\begin{subequations}\label{eq:OverallDecisions}
\begin{align}
\mbfb_{\text{CF}}=&[0~1~2~3~0~1~2~3]\\ 
\mbfb_{\text{IN}}=&[0~1~2~3~4~5~6~7],
\end{align}
\end{subequations}
respectively. As clearly seen in \eqref{eq:OverallDecisions}, the output labels of interference network are completely random. Whereas the output labels of $\mbcf$ network follow a full repetitive pattern, hence, they are significantly correlated. As demonstrated in Section \ref{sec:NumericalResults}, RFt with classifier chains is the key approach to exploit the correlated outputs to retain $99\text{-}100\%$ of the original $\mbsr$ results achieved by the proposed $\mbwc$ designs. 

\begin{figure}[!t]
\centering
\begin{subfigure}[t]{0.5\textwidth}
\centering
\includegraphics[scale=0.6] {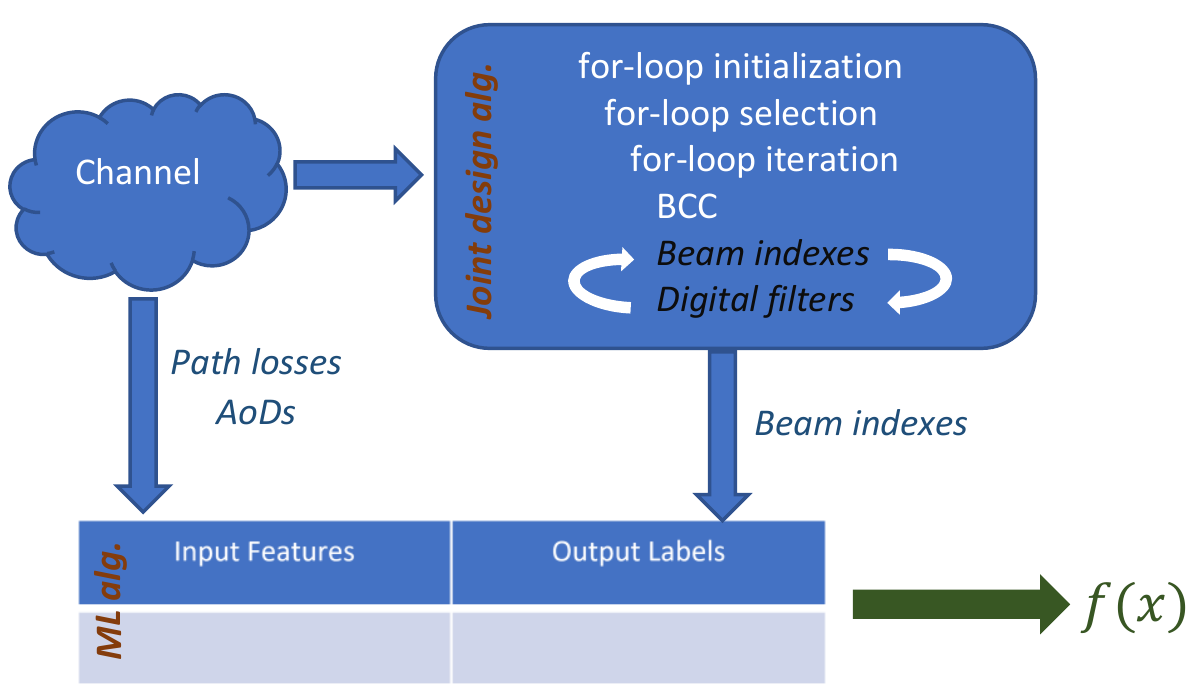}
\caption{Off-line ML training.}
\label{Fig:ImplicitDesignA_v01}
 \end{subfigure} 
\begin{subfigure}[t]{0.5\textwidth}
\centering
\includegraphics[scale=0.6] {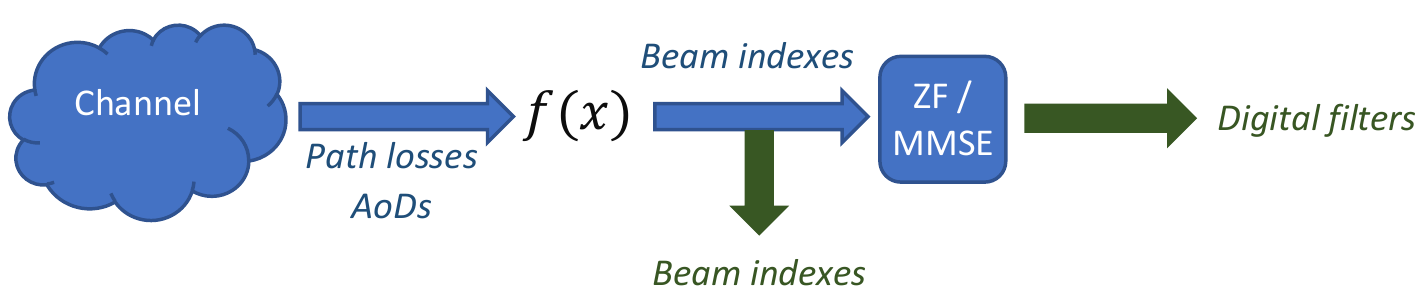}
\caption{Online ML testing followed by a digital filter design.}
\label{Fig:ImplicitDesignB_v01}
\end{subfigure}
\caption{Off-line training and online testing of ML algorithms.}
\label{Fig:ImplicitDesign}
\end{figure}

\subsection{Implicit Joint Design with Multiple Features}\label{subsec:Implicit}
\mbox{Off-line} training and online testing of ML algorithms are achieved as follows. For each of the training and testing instances, $2LK$ path loss and AoD values between the APs and users are input as feature vectors. For $\mbol$ training, $LK$ beam indexes between APs and users are used as output labels. The beam indexes can be determined by any of the joint design algorithms proposed in Section \ref{sec:BeamSelection}. Then, ML algorithms efficiently approximate the $\mbio$ mapping functions of the beam selection algorithms by using the feature vectors and output labels. For online ML testing, the trained ML algorithms are executed to determine the output labels, i.e., the beam indexes, by using the approximate mapping functions. At this stage, as detailed in Section \ref{subsec:DigitalPrecoder}, the effective channels are estimated so that a digital precoder, \eqref{eq:ZFFilter} or \eqref{eq:MMSEFilter} is designed in the last step. The approach of beam selection by an ML algorithm and then designing a digital precoder is implicitly a $\mbwc$ design since the ML algorithm is trained by a $\mbwc$ design. As illustrated in Figure \ref{Fig:ImplicitDesign}, both $\mbol$ training and online testing of ML algorithms are executed $\mbot$, i.e., there are no $\mbfls$. Moreover, the need of our proposed joint designs for frequent  effective channel estimations, as detailed in Section \ref{subsec:DigitalPrecoder}, are resolved by ML algorithms. For simplicity, in Figure \ref{Fig:ImplicitDesignA_v01}, one channel training instance is displayed, hence the $\mbio$ table has a single row.

\subsection{Scikit Implementation}\label{subsec:Scikit}
For the implementations of SVM, MLP, and RFt algorithms, Scikit library \cite{scikit} is used. These algorithms can be queried in \cite{scikit} by the following method names: LinearSVC and SVC (both belong to the sklearn.svm module), MLPClassifier (belongs to the sklearn.neural$\_$network module), and RandomForestClassifier (belongs to the sklearn.ensemble module). To access other classification methods under these modules, the module names can be queried as well, e.g., sklearn.ensemble can be searched in the top-right box \cite{scikit}.

As detailed earlier, $\mbmo$ classification problems can be solved by the naive binary relevance or the efficient classifier chain approach. These approaches can be queried by the wrappers MultiOutputClassifier and ClassifierChain, respectively, in \cite{scikit}. 

For hyperparameter optimization, GridSearchCV from $\mbdml$ library \cite{dask} is used for speeding the grid search method. In addition to $\mbdml$ library, there are many other alternative open source projects available online.

For the interested reader, the complete codes of the proposed beam selection algorithms are shared online \cite{YetisGitHubJointDesign}.

\begin{table}[!t]
\scriptsize 
\begin{center}
\caption{SOME ABBREVIATIONS AND NOTATIONS USED IN THE FIGURES.} \label{tab:Abbs}\vspace{-.2cm}
\begin{tabular}{|l|l|}
 \hline
(Corresponding figures)&\\
 $\tL x \tK y \tM z$ & A network with $L=x$ APs, $K=y$ users, \\
 &and $M=z$ antennas \\   
 init = $x$, iter = $y$ & Algorithm with number of initializations and iterations \\
 &are set to $x$ and $y$, respectively \\   
 (mm:ss) and (hh:mm) & The simulation durations in\\
 &(minutes:seconds) and (hours:minutes), respectively \\   \hline 
 (Figs. \ref{Fig:BeamSelec} and \ref{Fig:BeamSelectionAlg})&\\
  A - B - C&\\
 A: Semilinear / linear /  & Semilinear search algorithm (Section \ref{subsec:SemiLinearandSemiCentralizedSearches}) /   \\
 \hspace{.3cm}disjoint linear  & linear search algorithm (Section  \ref{subsec:LinearSearch}) / Naive disjoint\\   
   &  linear design. All other algorithms are joint designs. \\   
 B: II / IIS & Multiple initializations and iterations / \\  
 &multiple initializations, iterations, and selection  \\   
  C: Rate / DL & Selection based on the rate metric - equation \eqref{eq:RateSINR} /  \\
   &based on the direct link metric - equation  \eqref{eq:LinkPower} \\   \hline
(Fig. \ref{Fig:BCC})&\\
 w/ or w/o BCC & Algorithm with or without beam conflict control \\   
   &  process presented in Section  \ref{subsec:BCC} \\ 
 w/ BCC init & Algorithm with only beam conflict control  \\   
   & initializations, i.e., no beam conflict control process \\ \hline
(Fig. \ref{Fig:SMrf})&\\
  A - B - C&\\
 A: Linear, B: II  & Explained earlier in the table   \\
  C: Naive / Smart & The number of RF chains are naively fixed to a number / \\
   & are adaptively set to a number based on a threshold value \\   \hline   
\end{tabular}
\end{center}
\vspace{-.5cm}
\end{table}

\section{Numerical Results}\label{sec:NumericalResults}
In this section, we present the $\mbsr$ and simulation run time results for the proposed beam selection and ML algorithms. For all simulations, the channel center frequency and bandwidth are $28$ GHz and $850$ MHz, respectively, the path loss exponent is $2$, the $\mbln$ shadowing is $4$ dB, the AP transmit power is $43$ dBm, power spectral density of the white Gaussian noise is $-174$ dBm/Hz, and finally, the distances between APs and users vary uniformly between $95\text{-}105$ m. The channel gains $\beta\sikl^p$ are drawn i.i.d. from ${\mathcal{CN}(0,1)}$ distribution, and the number of paths, $P$, is assumed to be $1$.

DFT codebook can be adopted as a hardware friendly orthogonal beamforming solution. The element with the row and column indexes ${(x,y)}$, ${x,y=1,\ldots,M}$ of the DFT codebook matrix is given by 
\begin{equation}\label{eq:DFTCodebook}
\mbfD(x,y)=\frac{1}{\sqrt{M}}e^{-\frac{j2\pi(x-1)(y-1)}{M}},
\end{equation}
where ${j=\sqrt{-1}}$.
Note that the codeword $b$, i.e., $\mbflu_b$ in the codebook $\boldsymbol{\mclU}$ which is defined in Section \ref{subsec:ReceivedSignal}, is the $b\ssth$ column of $\mbfD$ in \eqref{eq:DFTCodebook}.

In Section \ref{subsec:ML}, the numerical results of ML algorithms with classifier chains are presented. In all cases tested, there is a sharp contrast in performance, as mentioned earlier, when classifier chains are not and are used, i.e., $63\%$ and $99\%$, respectively. Since including without classifier chain results in the figures can negatively impact the readability of the figures while not bringing beneficial information, they are not presented. The input features for ML algorithms are the sets of path losses and AoD. To avoid bias in the training phase, features are normalized before ML algorithms are executed. 

For all simulations, a desktop computer with Intel ${\text{i}7\text{-}8700}$ CPU, $3.20$ GHz, $16$ GB RAM, with $6$ cores and $12$ logical processors is used. For the beam selection and ML simulations, Matlab and Python with Scikit library are used, respectively. For the grid search tasks in ML algorithms, $\mbdml$ library is used. 

In Table \ref{tab:Abbs}, some abbreviations and notations used in the figures presented in the following sections are briefed.

\subsection{Beam Selection Results}

\begin{figure}[!t]
\centering
\begin{subfigure}[t]{0.5\textwidth}
   \includegraphics[scale=0.55] {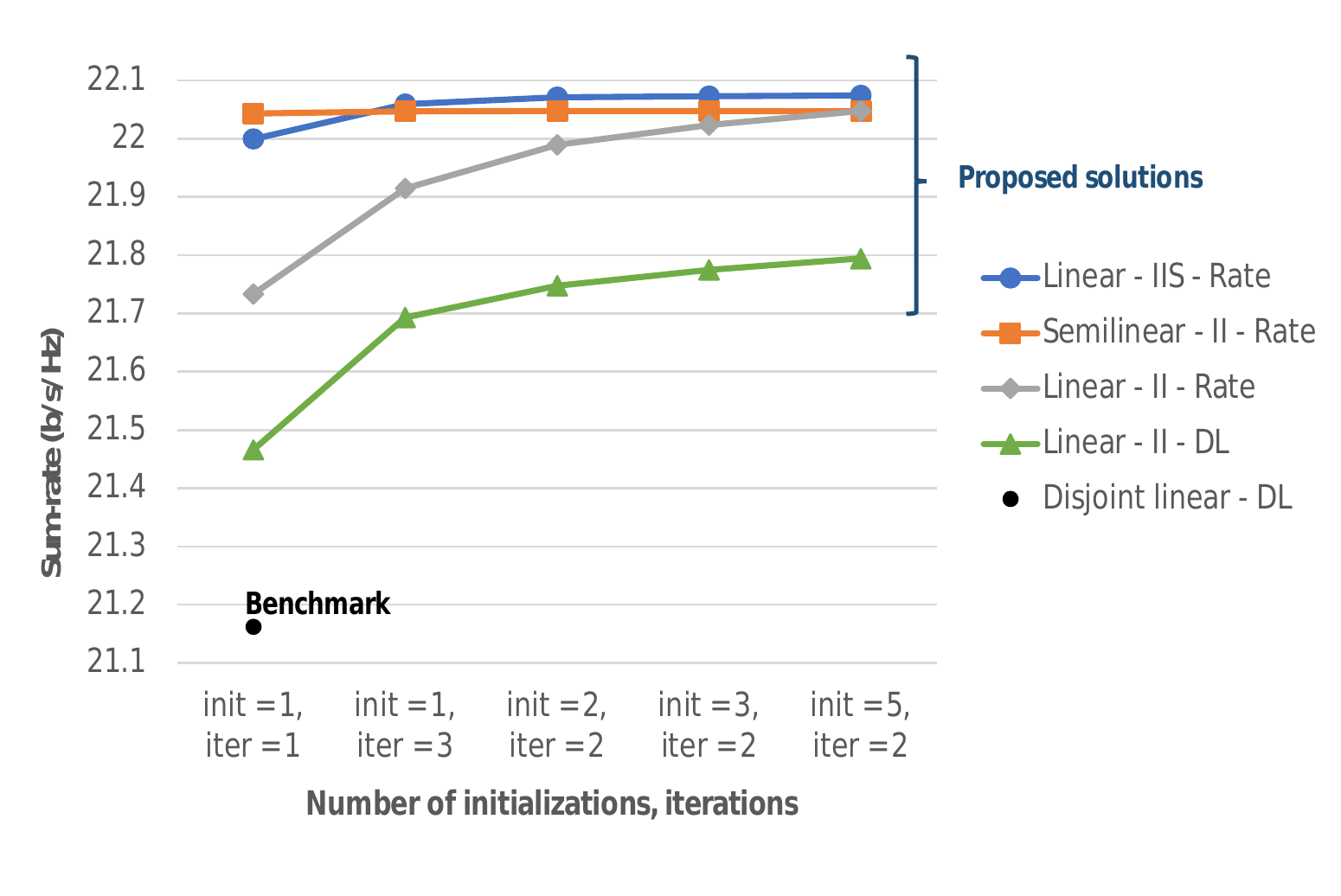}
   \vspace{-.5cm}
\caption{Sum-rates.}
\label{Fig:BeamSelecSumRateInitIter}
 \end{subfigure} 
 
\begin{subfigure}[t]{0.5\textwidth}
\centering
   \includegraphics[scale=0.55] {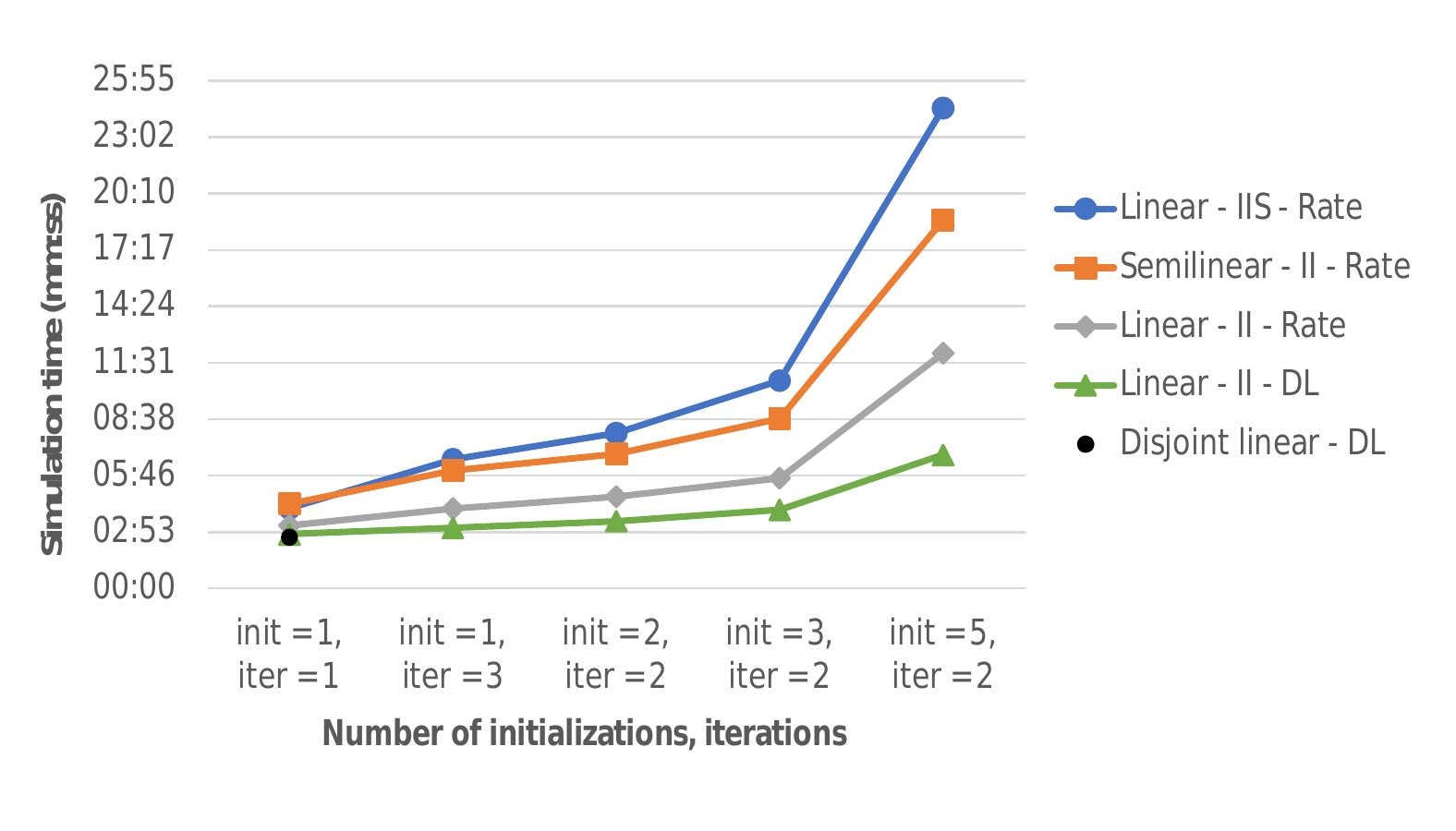}
   \vspace{-.5cm}
\caption{Simulation durations.}
\label{Fig:BeamSelecSimTimeInitIter}
\end{subfigure}
\caption{Sum-rates and simulation durations of the beam selection algorithms  with different numbers of initializations and iterations for the $\tL 3\tK 2\tM 8$ network.}
\label{Fig:BeamSelec}
\end{figure}

\begin{figure}[!t]
\centering
\begin{subfigure}[t]{0.5\textwidth}
\centering
  \includegraphics[scale=0.62] {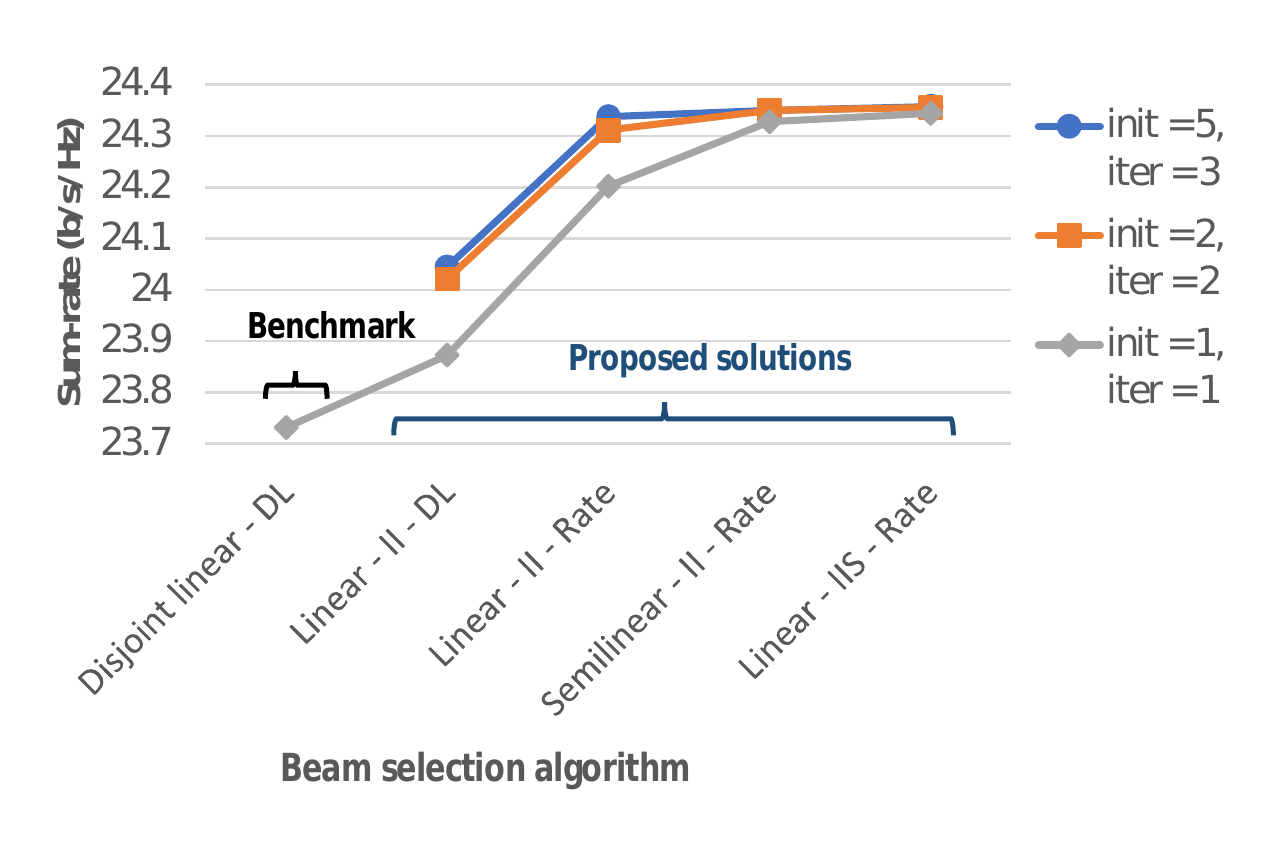}  
  \vspace{-.5cm}
\caption{Sum-rates.}  
\label{Fig:BeamSelectionAlgSumRate}
\end{subfigure}
 
\begin{subfigure}[!t]{0.5\textwidth}
\centering
    \includegraphics[scale=0.52] {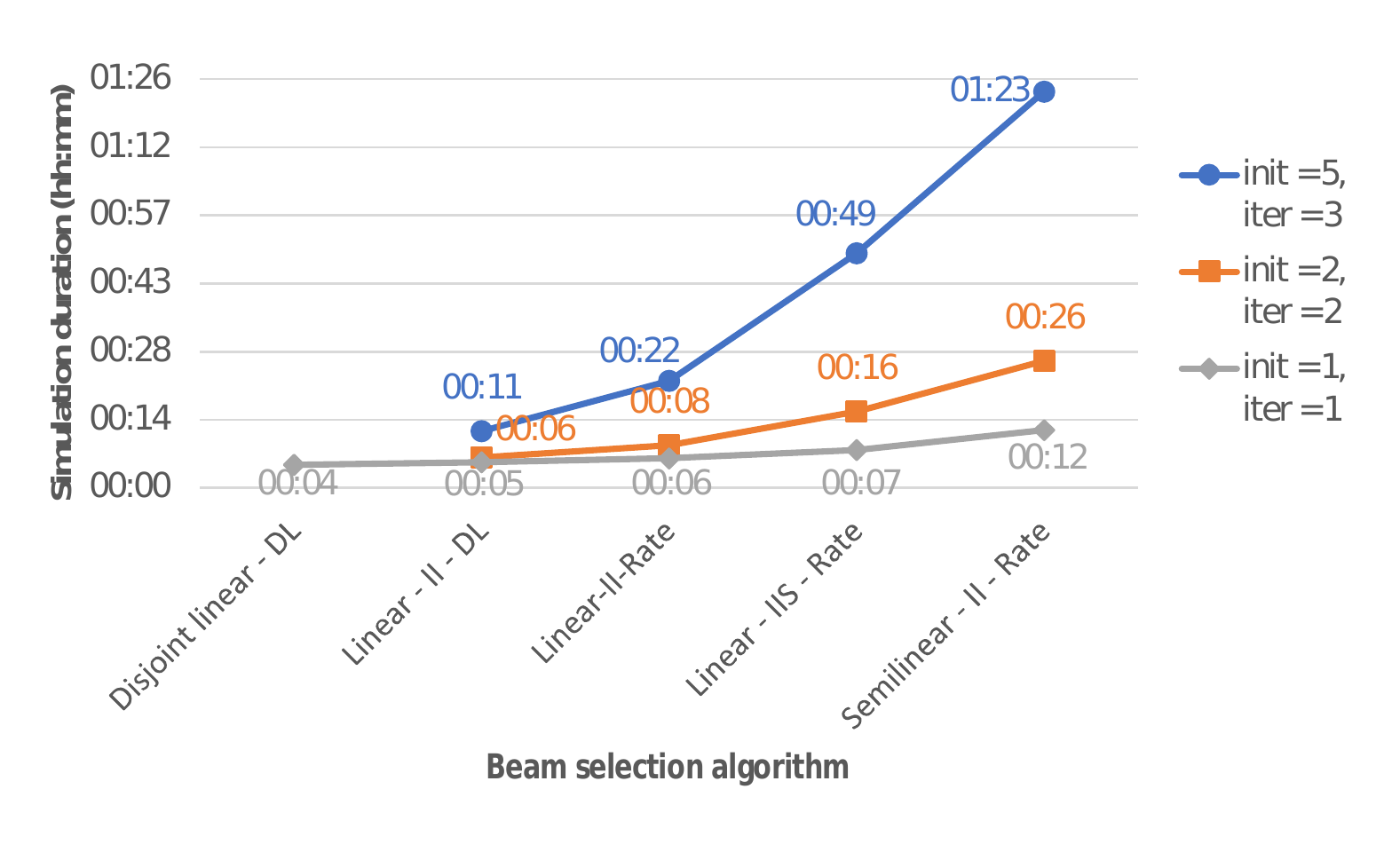}
    \vspace{-.5cm}
\caption{Simulation durations.} 
\label{Fig:BeamSelectionAlgSimTime}
\end{subfigure}
\caption{Sum-rates and simulation durations of beam selection algorithms with different numbers of initializations and iterations for the $\tL 3\tK 2\tM 16$ network.} 
\label{Fig:BeamSelectionAlg}
\end{figure}

In Fig. \ref{Fig:BeamSelec}, the $\mbsr$ results and the simulation run times of the beam selection algorithms proposed in Section \ref{sec:BeamSelection} are presented for different number of initializations and iterations. In the network, it is assumed there are ${L=3}$ APs, ${K=2}$ users, and ${M=8}$ antennas at each AP which can be denoted by $\tL L\tK K\tM M$, i.e., the network parameter notations are followed by their numerical values. The number of  Monte Carlo (MC) runs, i.e., the number of channel realizations, to obtain the numerical results is $5{\times10^3}~(5\tk)$. 

Clearly, as the beam selection algorithm becomes $\mbwc$ from a disjoint design to the complete proposed solution in Algorithm \ref{alg:SearchAlgs}, the $\mbsr$ results improve. However, as seen in Fig. \ref{Fig:BeamSelecSimTimeInitIter}, the improvement is marginal against the cost of increased simulation durations. As the numbers of initializations and iterations increase, the linear-II-rate algorithm achieves almost the same results as the linear-IIS-rate algorithm with a tolerable increase in the simulation duration. Whereas the $\mbsr$ gap to the disjoint linear-DL algorithm increases and the linear-II-DL algorithm still lags behind significantly.

In Fig. \ref{Fig:BeamSelectionAlg}, the same beam selection algorithms are tested in the $\tL 3\tK 2\tM 16$ network for again $5\tk$ MC runs. As the network size grows, the high complexity of $\mbslin$ algorithm starts to take effect and its simulation duration becomes the longest among all selection algorithms. linear-IIS-rate without multiple initializations and iterations achieves a remarkable $\mbsr$ result with a tolerable increase in the simulation duration. This result stresses the influence of the first search segment, i.e., AP, to do the beam selection on the $\mbsr$.

\begin{figure}[!t]
\centering
\begin{subfigure}[t]{0.5\textwidth}
  \includegraphics[scale=0.5] {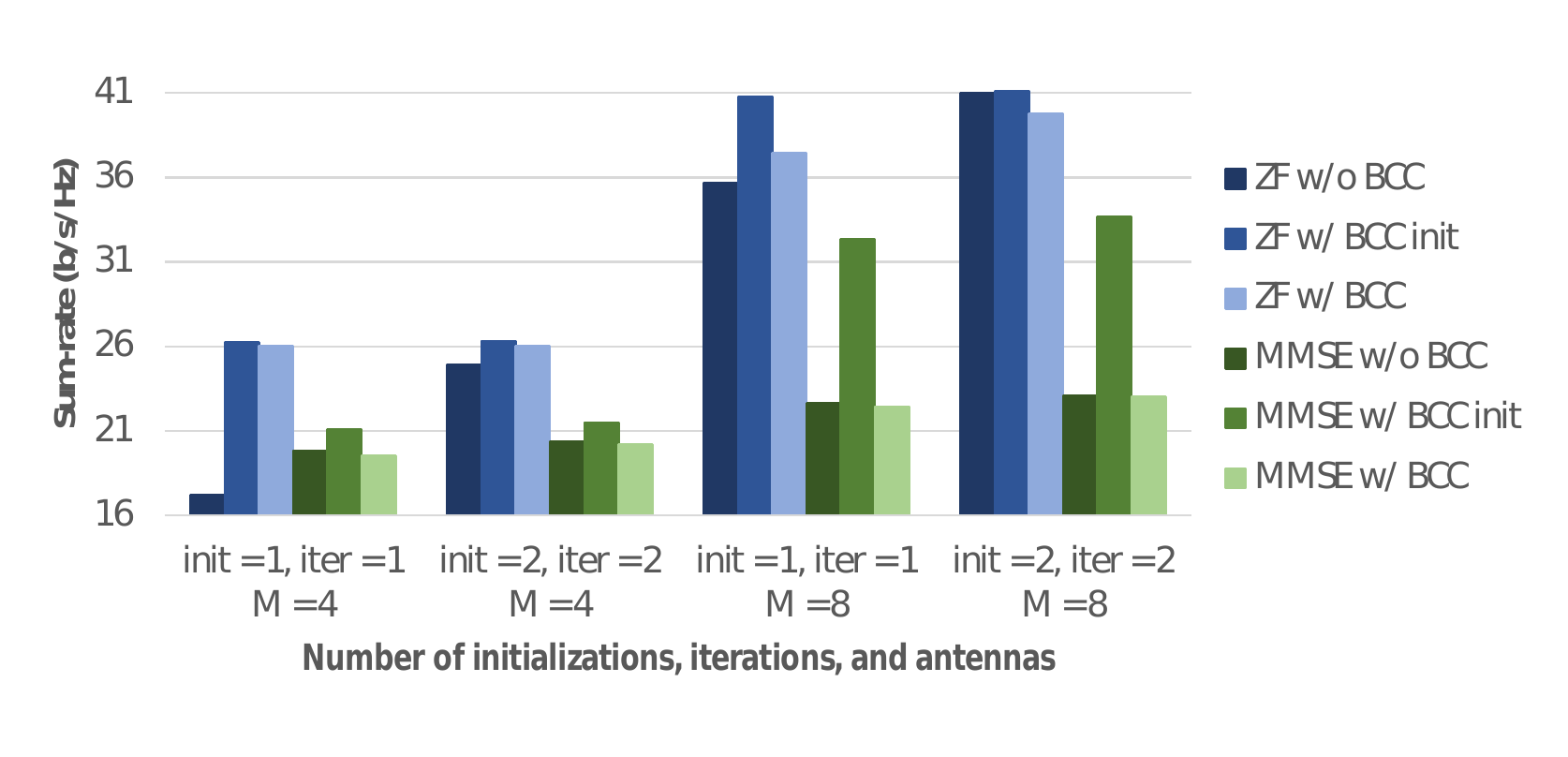}
    \vspace{-.5cm}
\caption{Sum-rates.}  
\label{Fig:BCC_SumRate}
 \end{subfigure} 
 \begin{subfigure}[t]{0.5\textwidth}
\includegraphics[scale=0.5] {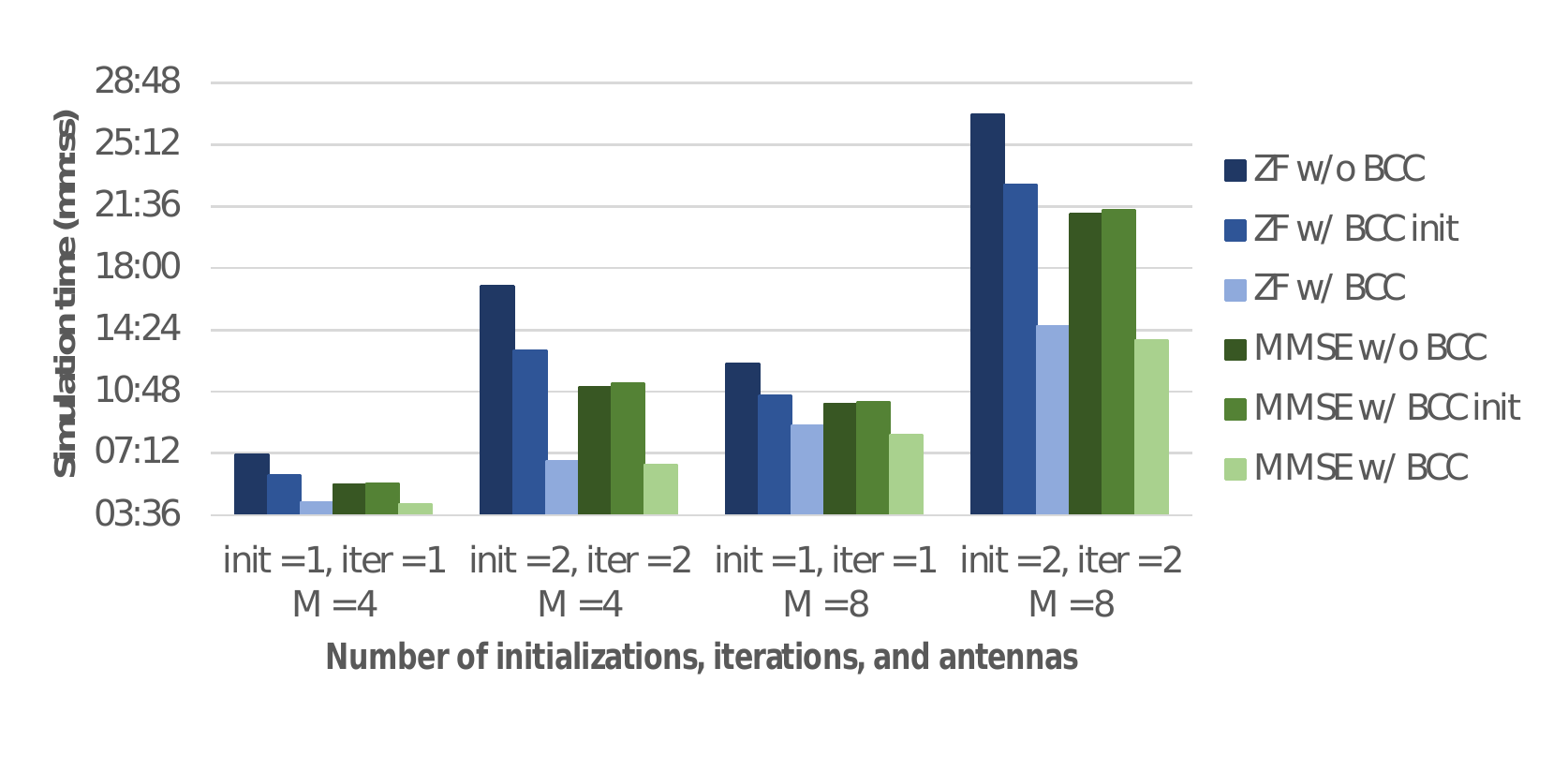}
  \vspace{-.5cm}
\caption{Simulation durations.}  
\label{Fig:BCC_Time}
 \end{subfigure} 
 \caption{Sum-rates and simulation durations of the linear-II-rate algorithm with BCC, without BCC, and with BCC initialization for the $\tL 4\tK 4\tM 4$ and $\tL 4\tK 4\tM 8$ networks.}  
\label{Fig:BCC}
\end{figure}

In Fig. \ref{Fig:BCC}, the effects of BCC on the $\mbsr$ and simulation duration are demonstrated over varying number of initializations, iterations and antennas. For all results, linear-II-rate beam selection algorithm is used. Algorithms with and without the BCC implementation are denoted by w/ BCC and w/o BCC, respectively. For the algorithms without the BCC implementation, we also propose random initializations that satisfy BCC, i.e., BCC initializations. These algorithms are denoted by w/ BCC Init. In particular, for the w/ BCC Init algorithms, BCC condition for random initialization is active in step \ref{step:SearchAlgsInit} of Algorithm \ref{alg:SearchAlgs} but the BCC implementation in step \ref{step:BCCLinear} of Algorithm \ref{alg:SearchAlgsB} is inactive. As detailed next, our numerical results indicate that the three options, i.e., the algorithms with and without the BCC implementation, and with the BCC initializations, offer \mbox{trade-offs} between the $\mbsrs$ and simulation durations.

When there is no BCC implementation, the effective channel in \eqref{eq:EffectiveChannel} can be $\mblr$ and can have a high, i.e., poor, channel condition number. As seen in Fig. \ref{Fig:BCC_SumRate}, ZF with the BCC implementation achieves a higher $\mbsr$ than ZF without the BCC implementation until a point, i.e., ${\text{init}=2\text{, iter}=2, M=8}$. BCC implementation reduces the number of beam combinations. Thus, BCC implementation is significantly effective in reducing the simulation duration as seen in Fig. \ref{Fig:BCC_Time}. However, BCC implementation also reduces the search space so that the point that achieves a higher $\mbsr$ can be excluded in the search process. In Fig. \ref{Fig:BCC_SumRate}, from left to right along the $\mbox{\text{x-axis}}$, the search space increases since the numbers of random initializations and antennas are increasing. Hence, the advantage of larger search space becomes dominant over the advantages of higher channel rank and lower channel condition number. MMSE without the BCC implementation can achieve a higher $\mbsr$ than MMSE with the BCC implementation due to the increased search space and increased robustness of MMSE against the $\mblr$ and high channel condition number of the effective channel. In all cases, with the BCC implementation option reduces the simulation duration compared to without the BCC implementation option. The reduction is more distinct when multiple random initializations are used.

For the algorithms without the BCC implementation, BCC initialization can increase the $\mbsr$ and decrease the simulation duration in general. In essence, this option benefits from the improved rank and channel condition number due to the BCC initialization, and also from increased search space although BCC implementation is not applied during the search processes. For both ZF and MMSE filters, BCC initialization achieves the highest $\mbsrs$. For ${\text{init}=1\text{, iter}=1, M=8}$, BCC initialization significantly improves the $\mbsrs$ of the ZF and MMSE filters, respectively, compared to ZF and MMSE without the BCC implementation. The simulation duration of ZF with BCC initialization is lower compared to ZF without BCC. However, for the MMSE filter, BCC initialization slightly increases the simulation duration compared to MMSE without BCC. For the MMSE filter, it is observed that although BCC initialization increases the rank, it also increases the channel condition number. Increased channel condition number causes numerical instability thus the simulation duration is increased. 

\begin{figure}[!t]
\centering
  \includegraphics[scale=0.55] {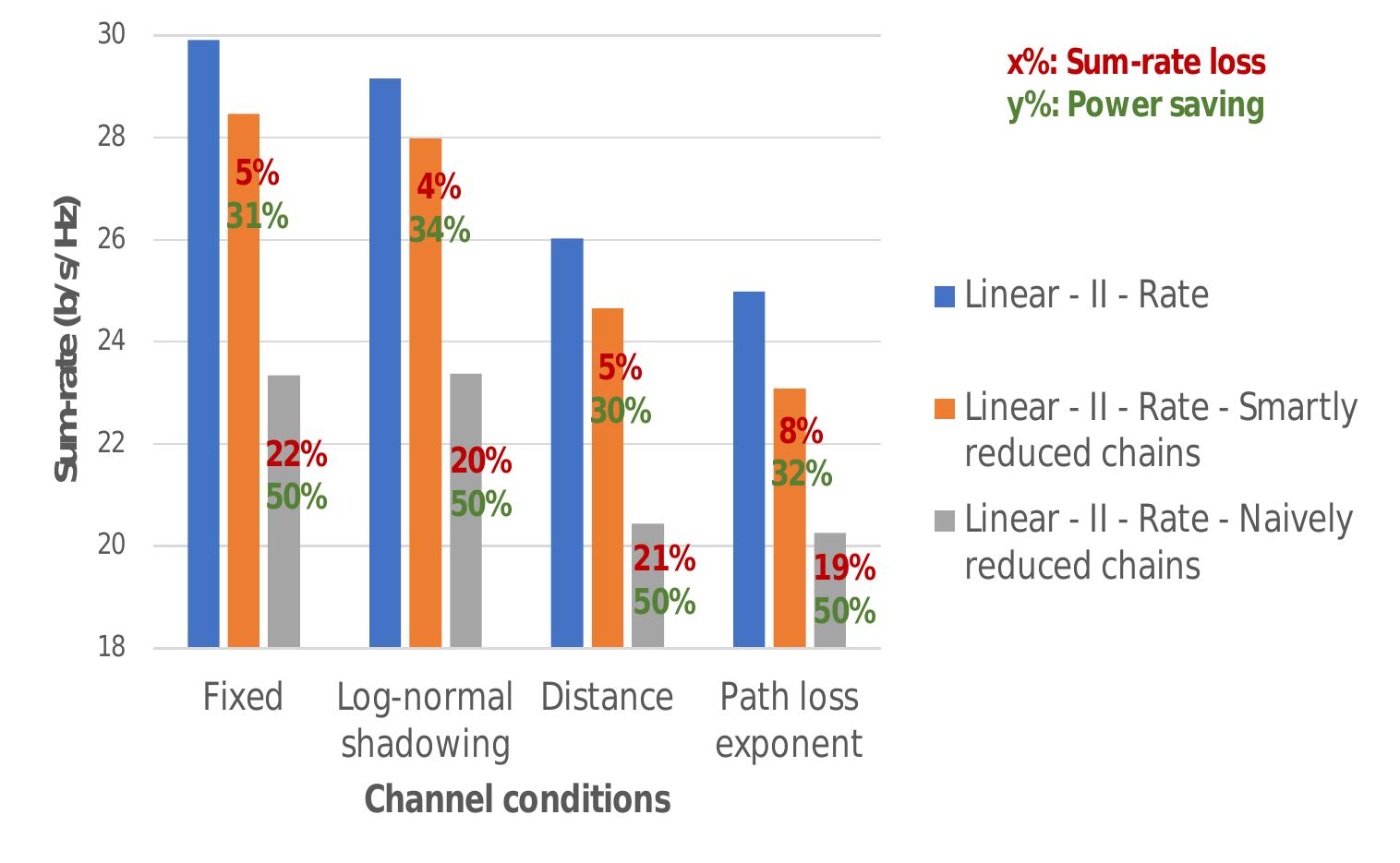}
\caption{Sum-rates of the linear beam search algorithm in different channel conditions for the $\tL 4\tK 4\tM 4$ network when the numbers of RF chains are full (${M_\trf=K}$), smartly reduced (${M_\trf<K}$ statistically), and naively reduced (${M_\trf<K}$ deterministically).}  
\label{Fig:SMrf}
\vspace{-.3cm}
\end{figure}

In Fig. \ref{Fig:SMrf}, the $\mbsrs$ are evaluated for varying channel conditions. For all results, linear-II-rate with a single initialization and iteration beam selection algorithm is used. As the channel conditions get poor, the marginal benefit of serving all users can be low considering the $\mbsr$ gain versus the power cost. In poor channel conditions, shutting off some RF chains can lower the power consumption at a low $\mbsr$ loss. For the naive approach in Fig. \ref{Fig:SMrf}, denoted by linear-II-rate-naive, the number of RF chains are reduced to a fixed number, ${M_{\trf}=2}$. On the other hand, for the smart approach, denoted by linear-II-rate-smart, the number of RF chains are smartly reduced based on a threshold value determined by the path loss value in \eqref{eq:PathLoss}. In particular, the threshold is set to the mean minus $1/4$ power of variance of the path losses between AP $l$ and all users. Users with the path losses below this threshold are served, and the others are not, thus the corresponding RF chains of those users are shut off. The blue bars in Fig. \ref{Fig:SMrf} indicate the case when all RF chains are in use, i.e., ${M_\trf=4}$. The \mbox{left-most} set of results are obtained when channel conditions are fixed to the certain values given in the beginning of this section. The $2\ssnd$ and the following set of results from left are obtained when $\mbln$ shadowing, distance, and path loss exponent vary with uniform distribution between \mbox{$4$-$6$} dB, \mbox{$100$-$200$} m, and \mbox{$2$-$4$}, respectively. The bold red and green numbers inside the bars are the $\mbsr$ loss and power saving percentages when compared to the blue bars. The power savings for the naive cases are always $50\%$, i.e., $M_\trf$ is set to $2$ instead of $4$. The results clearly indicate that by smartly shutting off the RF chains, significant power savings can be gained at a cost of small $\mbsr$ losses. Similar to the selection feature explained earlier, smartly reducing the chains can be more advantageous in asymmetric networks while raising the fairness issue in the network.

As also demonstrated in Fig. \ref{Fig:SMrf}, the benchmarks of the considered algorithms in our work are indifferent to varying channel conditions. Over varying channel conditions, the $\mbsr$ losses of smart and naive approaches vary around $5\%$ and $20\%$ in average, respectively, and the power saving of smart approach varies around $30\%$ in average. Only the benchmarks of the selection and smartly reduced chains features are expected to be different in asymmetric versus symmetric channels as mentioned earlier.

\begin{figure}[!t]
\centering
  \includegraphics[scale=0.55] {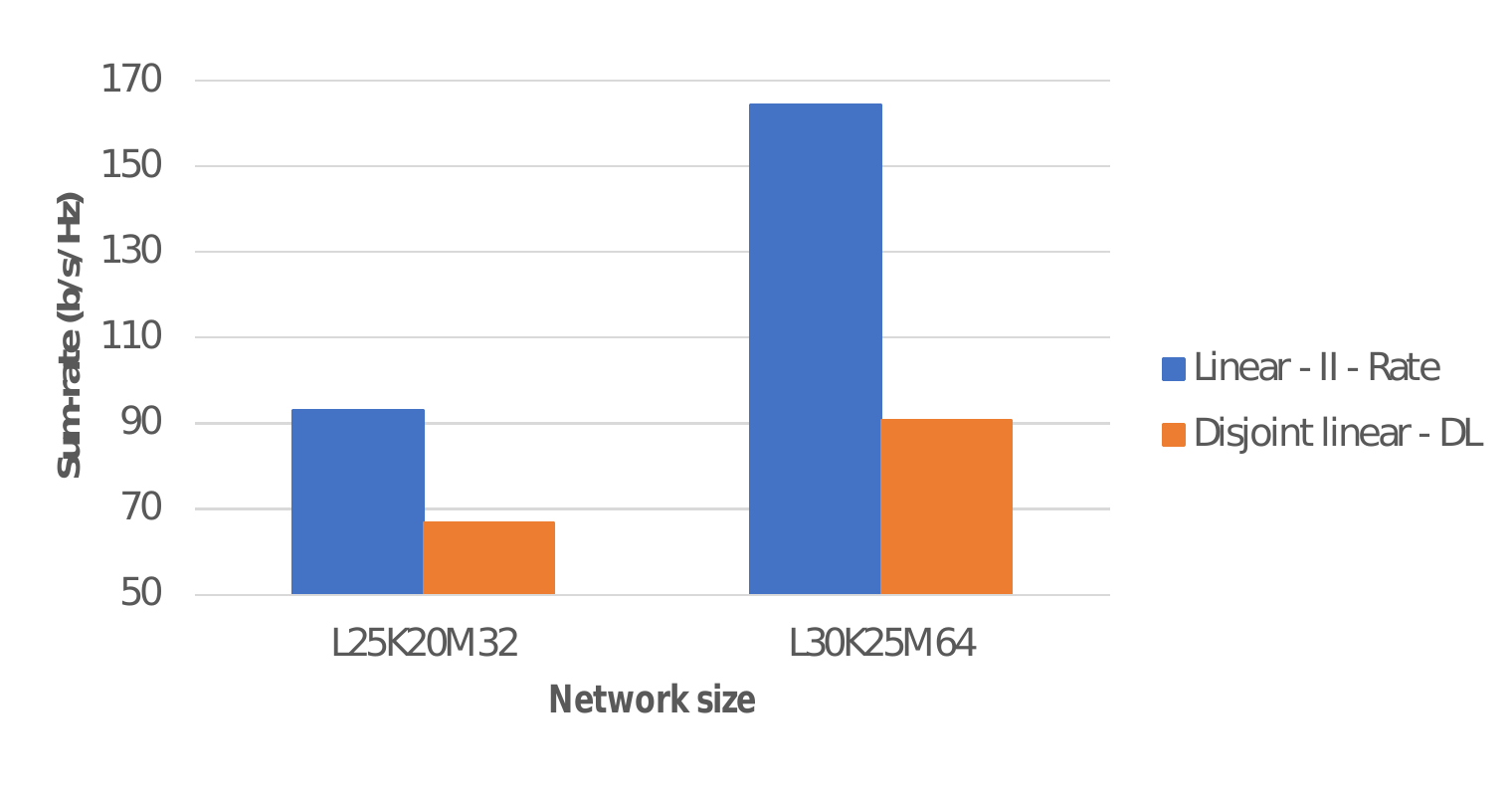}
\caption{Sum-rates of the joint and disjoint linear beam selection algorithms for large network sizes. For the joint design, ${\text{init}=1\text{ and iter}=1}$.}  
\label{Fig:LargeNetwork}
\vspace{-.2cm}
\end{figure}

Our proposed solutions can substantially benefit from larger network sizes as detailed next. In Fig. \ref{Fig:LargeNetwork}, the sum-rates are evaluated for large network sizes. For the results, the path loss exponent, $\mbln$ shadowing, and the distances between APs and users vary uniformly between $2\text{-}8$, $4\text{-}10$ dB, and $10\text{-}200$ m, respectively. Due to the large network sizes, MC run is set to $100$. For linear-II-rate, single initialization and iteration are used. As the network size grows, same conclusions presented earlier in this section can be drawn while the numerical gaps, i.e., benchmarks, between the results increase. As seen in Fig. \ref{Fig:LargeNetwork}, the advantage of joint design becomes significant over the naive disjoint design as the network size grows. As the network size grows, more complex algorithms are likely to benefit more than the simpler algorithms by using the rich degrees of freedom in larger networks. 

The complexity of the linear search algorithm for the largest network size presented earlier in Fig. \ref{Fig:BCC}, i.e., $\tL 4\tK 4\tM 8$, is only 128. On the other hand, the complexities of the networks $\tL 25\tK 20\tM 32$ and $\tL 30\tK 25\tM 64$ in Fig. \ref{Fig:LargeNetwork} are $16\tk$ and $48\tk$, respectively. Even for small network sizes, the $\mbslin$ algorithm presented in Section \ref{subsec:SemiLinearandSemiCentralizedSearches} becomes impractical. For the networks $\tL 4\tK 3\tM 16$ and $\tL 4\tK 4\tM 8$, the complexity of the $\mbslin$ algorithm is nearly $16\tk$. Interested reader can refer to the spreadsheet in \cite{YetisGitHubJointDesign} for the complexities of varying network sizes. As detailed in Section \ref{sec:BeamSelection}, the complexity evaluations are based on the searching principles without the extra features. If the multiple initializations, iterations, and the selection features (IIS) are included in an algorithm, its complexity value is multiplied by init$\times$iter$\times L$.  

Next, we benchmark the $\mbsrs$ and run times of ML algorithms. 

\subsection{Machine Learning Results}\label{subsec:ML}

\begin{figure}[!t]
\centering
  \includegraphics[scale=0.55] {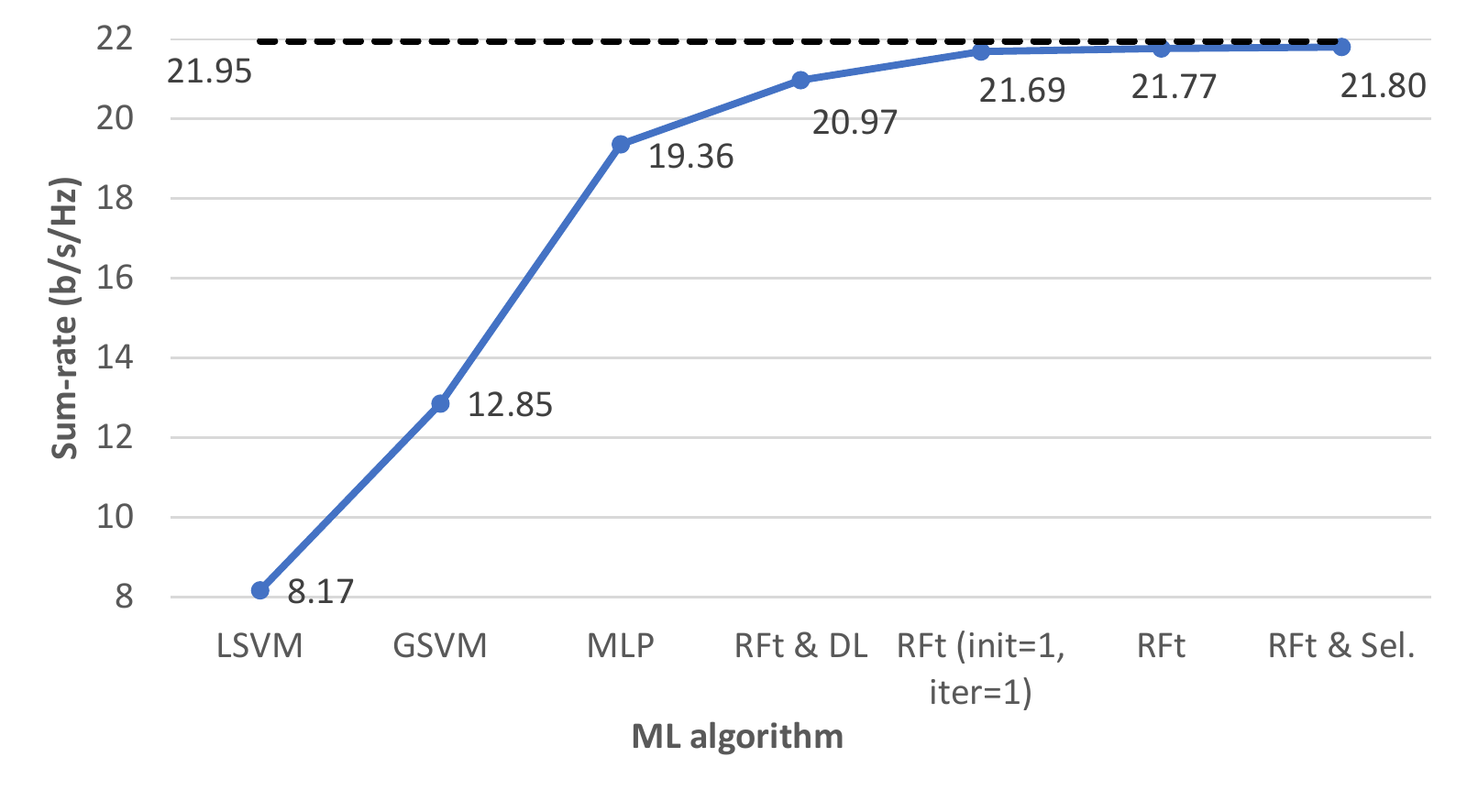}
\caption{Sum-rates of ML algorithms for the $\tL 3\tK 2\tM 8$ network. RFt $\&$ DL, RFt $\&$ Sel., and the rest are trained by the linear-II-DL, linear-IIS-rate, and linear-II-rate algorithms. Unless noted, init $=2$, iter $=2$ parameters are set.}  
\label{Fig:MLAlgSumRate}
\vspace{-.2cm}
\end{figure}

In Fig. \ref{Fig:MLAlgSumRate}, the $\mbsr$ results of ML algorithms for the $\tL 3\tK 2\tM 8$ network are presented. For the training and testing phases, $30\tk$ and $10\tk$ instances are executed, which give a $75\%$ to $25\%$ ratio of all data, respectively. The training and test data are independently generated. Nevertheless, all ML algorithms are passed through ${5\text{-fold}}$ $\mbcv$ (CV), and also, through hyperparameter optimization stages.

To train and test the ML algorithms, based on the numerical results presented earlier, the linear-II-rate algorithm with $2$ initializations and iterations (i.e., init $=2$, iter $=2$) is chosen since it can achieve $\mbsrs$ close the best results presented earlier within short simulation durations. This choice is primarily based on saving time for the increased durations of training and testing of ML algorithms and note that the benchmarking of ML algorithms is independent of beam selection algorithms. 

For the training of the linear SVM (LSVM), Gaussian (GSVM), MLP, and RFt algorithms, linear-II-rate beam selection algorithm is used. Except RFt with init=1 and iter=1, the numbers of initializations and iterations are set to $2$. For the training of RFt$\&$DL and RFt$\&$Sel., linear-II-DL and linear-IIS-rate beam selection algorithms are used, respectively. 

In Fig. \ref{Fig:MLAlgSumRate}, the dashed line on top is the $\mbsr$ obtained by the proposed linear-IIS-rate beam selection algorithm, i.e., the maximum original $\mbsr$ that can be achieved by the ML algorithms. Clearly, upon with the utilization of the RFt algorithm, $\mbsrs$ between $96\text{-}99\%$ of the original results can be achieved. 

As noted earlier, Matlab and Python languages are used for the beam selection and\ ML algorithms. Nevertheless, we can conclude that RFt can achieve remarkable $\mbsr$ results within tolerable durations compared to conventional beam selection algorithms as follows. The total simulation durations of  training and testing phases for RFt, MLP, GSVM, and LSVM algorithms  are about 01:30, 01:00, 15:00, and 00:05 (mm:ss), respectively. Hence, we can conclude that the RFt algorithm has the best accuracy with a reasonable speed. The simulation duration for the dashed line in Fig. \ref{Fig:MLAlgSumRate} which is produced via the linear-IIS-rate beam selection algorithm  is about 12:00. Indeed, even more $\mbwc$ algorithms that have longer durations than the linear-IIS-rate algorithm are admissible to achieve higher $\mbsrs$ since the  beam selection algorithms are executed $\mbol$ to generate the   data that train the ML algorithms, and the simulation durations of the ML algorithms are unaffected by the training beam selection algorithms. The results in Figs. \ref{Fig:LargeNetwork} and \ref{Fig:MLAlgSumRate} clearly advocate that the more $\mbwc$ algorithm is used for ML training, the more advantage over naive disjoint design can be achieved in large network sizes.

\begin{figure}[!t]
\centering
  \includegraphics[scale=0.55] {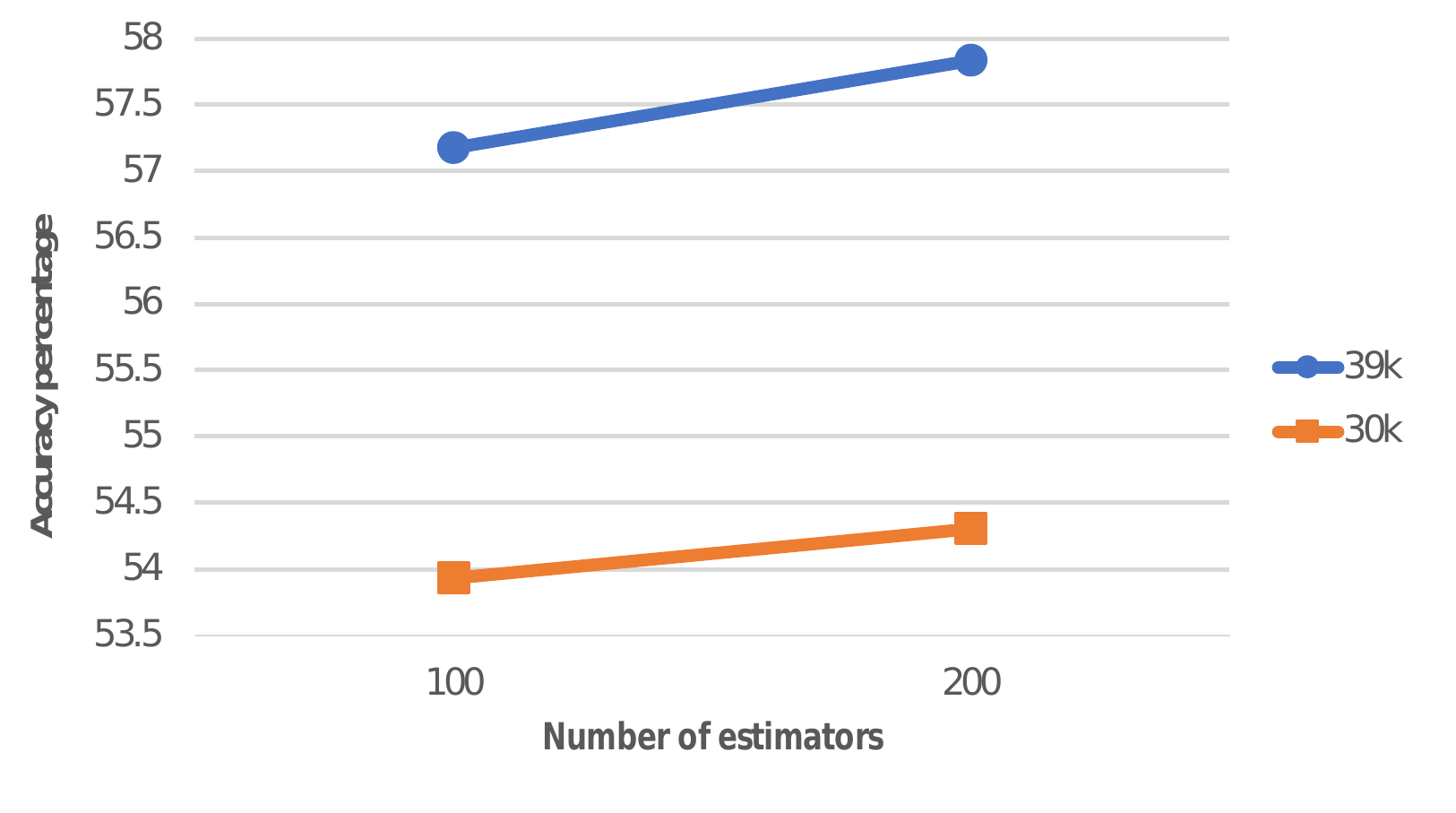}  
  \caption{The performance plots of accuracy percentage versus number of training data and number of estimators, i.e., trees, for the RFt classification algorithm in the $\tL3\tK2\tM8$ network. RFt is trained by the linear-II-rate algorithm.}
\label{Fig:IncreasingMC}
\vspace{-.2cm}
 \end{figure}

The classification accuracies of all ML algorithms can be improved by increasing the number training data as seen in Fig. \ref{Fig:IncreasingMC}, where the accuracy percentage of RFt is shown to increase as the training data is increased from $30$k to $39$k. Increasing the number of estimators, i.e., trees, in the case of RFt can also significantly improve the classification accuracy of the algorithm. Higher classification accuracies of ML algorithms result in higher $\mbsrs$ in return.

Among the ML algorithms, the highest accuracy is achieved by RFt$\&$DL which is $96\%$. Other RFt algorithms that use $\mbsr$ as the beam selection metric achieves about $55\%$ accuracy. The accuracies of MLP, GSVM, and SVM are about $45\%,25\%$, and $20\%$, respectively. The reason for high accuracy of RFt$\&$DL is that the input features for the ML algorithms are the sets of path losses and AoD. Thus, the DL power, as the beam selection metric matches better to these input features. Although the accuracy of RFt$\&$DL is high, linear-II-DL beam selection algorithm has a lower $\mbsr$, as seen in Figs. \ref{Fig:BeamSelecSumRateInitIter} and \ref{Fig:BeamSelectionAlgSumRate}, compared to the other selection algorithms that use $\mbsr$ as the beam selection metric. Thus, with  $55\%$ accuracy, RFt algorithms with the $\mbsr$ beam selection metric can achieve higher $\mbsr$ results as seen in Fig. \ref{Fig:MLAlgSumRate}. Our numerical results indicate that RFt algorithms with the $\mbsr$ beam selection metric can recognize alternative beam combinations that achieve high $\mbsr$ results although the selected beam combination by the RFt algorithm may not match to the beam combination that is originally chosen by the beam selection algorithm. When ${M=K}$, RFt algorithm can retain the original $\mbsr$ by $100\%$ although its accuracy is still unremarkable because, again, the RFt algorithm can identify alternative beam combinations different than the combinations chosen originally by the beam selection algorithm so that RFt algorithm can still achieve high $\mbsr$ results. In fact, if the exhaustive search algorithm, i.e., centralized algorithm, can be used instead, then the RFt algorithm with a $55\%$ accuracy rate may not retain $99\%$ of the original $\mbsr$ achieved by the exhaustive algorithm since the beam combination selections by the exhaustive search algorithm are optimum. However, as mentioned earlier, the exhaustive search algorithm is impractical due to its high complexity.

\begin{figure}[!t]
\centering
  \includegraphics[scale=0.65] {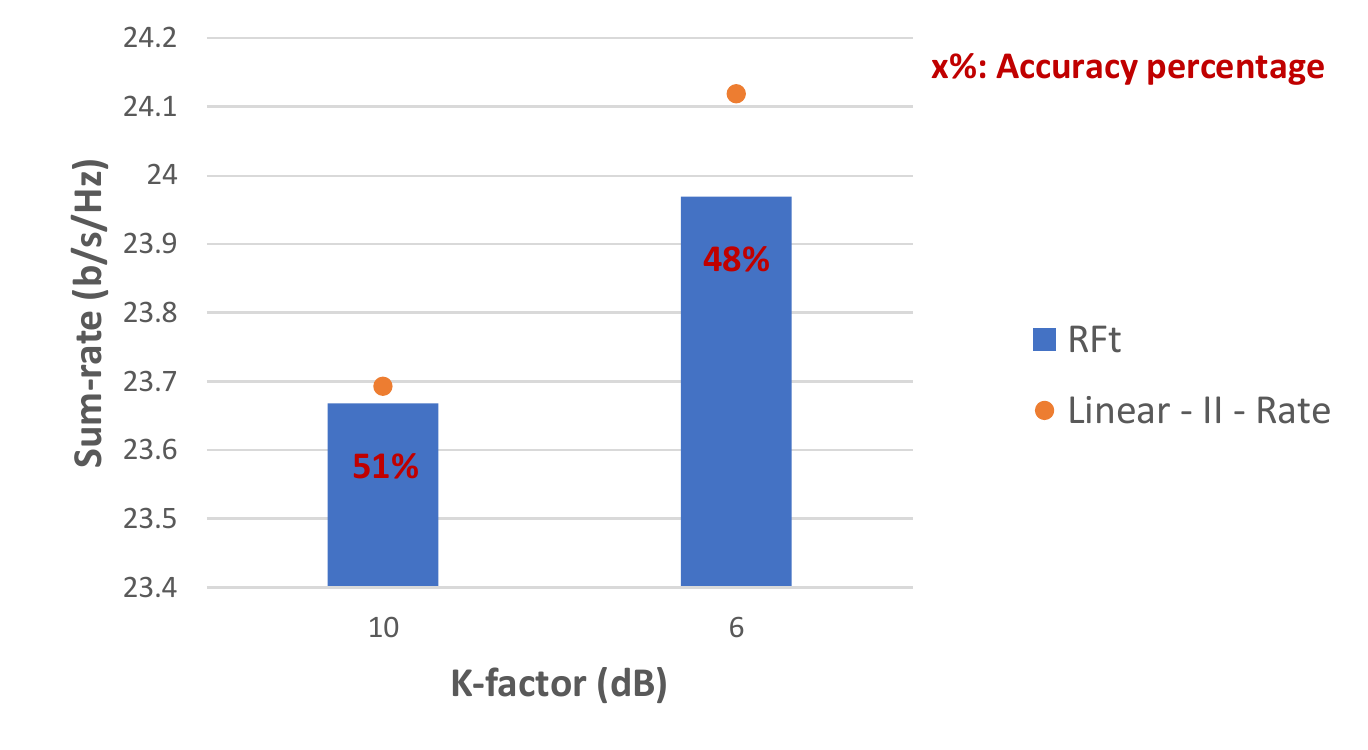}  
  \caption{Sum-rates and accuracy percentages over varying $\mbkf$ values for the RFt and linear-II-rate algorithms in the $\mbmp$ ($P= 3$) $\tL3\tK2\tM8$ network.}
\label{Fig:KFactor}
\vspace{-.2cm}
 \end{figure}

As demonstrated in Fig. \ref{Fig:KFactor}, the proposed joint designs and ML algorithms can be effective in $\mbmp$ $\mbmmw$ channels which typically have one strong $\mblos$ (LoS) path and a few reflected paths. A $\mbkf$ of $10$ dB implies that the LoS path is nearly $10$ dB stronger than the combined power of all reflected paths. Hence, analog beamforming can be effective by aligning the beam with the strong LoS path and ignoring the remaining paths.  Similarly, ML algorithms can be effective when the remaining paths are ignored. By ignoring the remaining paths, firstly, the length of input feature vector is reduced by $P$, e.g., instead of using $LKP$ path losses and AoDs, $LK$ of them are used as input features. Secondly, the data imbalance problem is avoided, for instance, when the absolute channel gains ${|\beta\sikl^p|}$ and path losses are used as input features. In Fig. \ref{Fig:KFactor}, the $\mbsr$ and accuracy percentage results of linear-II-rate (init $=1$, iter $=1$, and MMSE digital precoder) and RFt ($200$ estimators) algorithms are presented for the $\tL3\tK2\tM8$ network where ${P=3}$ is assumed. For the RFt algorithm, the features regarding  the strong LoS paths are considered. For both $\mbkf$ values $10$ and $6$ dBs, the RFt algorithm achieves almost the same $\mbsr$ results as the linear-II-rate algorithm although only the features of strong LoS paths are considered. This implies that the RFt algorithm can effectively identify the alternative solutions as explained earlier in Fig. \ref{Fig:MLAlgSumRate}.  On the other hand, the accuracy performance of RFt is challenged at low $\mbkf$ values. Thus, more MC runs and input features are needed for the training and testing phases of ML algorithms. For the results in Fig. \ref{Fig:KFactor}, the training and test instances are again set to $30\tk$ and $10\tk$, respectively. But, the absolute channel gains ${|\beta\sikl^p|}$ of the strong LoS paths are included in the feature vectors in addition to the path losses and AoDs for improving the accuracy percentages. Nevertheless, compared to the accuracy percentages presented in Fig. \ref{Fig:IncreasingMC} where $P=1$ is assumed, the achieved accuracy percentages are lower in Fig. \ref{Fig:KFactor}. This implies that the number MC runs need to be tuned based on the $\mbkf$ value to achieve the target accuracy percentages. In short, as demonstrated in this section, for ML algorithms, tuning the number of MC runs and number of input feature types  (e.g., absolute channel gains, path losses, and AoDs) depending on the $\mbkf$ value while considering only the strong LoS paths can be an effective solution. The proposed linear-II-rate algorithm can  be effectively implemented in $\mbmp$ channels as demonstrated in Fig. \ref{Fig:KFactor}. However, there can be situations where the reflected paths can impact the analog beamforming design. These situations can be addressed in a future work. 

\vspace{-.45cm}
\section{Conclusion}\label{sec:Conclusion}
In this work, we propose joint design algorithms of analog beam selection (based on the $\mbsr$ metric) and digital precoders. The joint designs are $\mbwe$ with multiple initializations, iterations, and selection features as well as with the BCC implementation. Hence, the network $\mbsr$ gains are much higher compared to the naive disjoint design of analog beam selection (based on the DL power metric) and digital precoders. We show that BCC initialized algorithms without the BCC implementation can achieve the best network $\mbsrs$ compared to the algorithms with and without the BCC implementation, and also, BCC initialization can reduce the simulation durations. Finally, in poor channel conditions, RF chains can be selectively shut off to save significant power consumptions at the cost of low network $\mbsr$ losses.
Next, we propose supervised ML algorithms that are trained by the beam selection decisions obtained from the $\mbwc$ algorithms. The numerical results obtained via the RFt algorithm are promising since it can retrain $99\text{-}100\%$ of the original $\mbsr$ results achieved by the proposed joint design algorithms.

Selectively shutting off the RF chains can only save power consumptions. As a future endeavour, further savings, e.g.,  chip area, can be achieved by serving more users than the number of RF chains. 
\vspace{-.45cm}
\bibliographystyle{IEEEtran}

\end{document}